\documentclass{aa}

\usepackage{graphicx}
\usepackage{natbib}

\usepackage[LGR,T1]{fontenc}

\usepackage{amsmath}
\usepackage{amssymb}
\usepackage{mathrsfs}
\usepackage{textcomp}
\usepackage{soul}

\usepackage{tikz}

\usepackage{float}

\usepackage{geometry}

\usepackage{subfig}

\usepackage{color}

\usepackage{hyperref}
\hypersetup{colorlinks=true,citecolor=blue,linkcolor=blue}

\usepackage{enumerate}

\newcommand{\Ham}{\mathcal{H}}

\newcommand{\drond}[2]{\frac{\partial #1}{\partial #2}}
\newcommand{\vect}[1]{\mathbf{\boldsymbol{#1}}}
\newcommand{\accolade}{\phantom{\frac{1}{1}\!\!\!\!\!}}

\newcommand{\R}{\mathcal{R}}
\newcommand{\place}{\;\;\;\;\;}

\newcommand{\LGR}{\fontencoding{LGR}\selectfont}
\newcommand{\Latin}{\fontencoding{\encodingdefault}\selectfont}

\newcommand{\stigma}{\text{\LGR \textstigma{} \Latin}\!\!}

\newcommand{\qoppa}{\text{\LGR \textqoppa{} \Latin}\!\!}

\newcommand{\be}{\begin{equation}}
\newcommand{\ee}{\end{equation}}

\newcommand{\gO}[1]{\mathcal{O}\left(#1\right)}

\newcommand{\Xb}{\bar{X}}

\graphicspath{{./image/}}
\newcommand{\bibpath}{../}

\begin{document}

\title{Dynamics of co-orbital exoplanets in a first order resonance chain with tidal dissipation}
\author{J\'er\'emy Couturier\inst{1}\and Philippe Robutel\inst{1}\and Alexandre C.M. Correia\inst{2,1}}
\institute{IMCCE, UMR8028 CNRS, Observatoire de Paris, PSL Univ., Sorbonne Univ., 77 av. Denfert-Rochereau, 75014 Paris, France
	\and
	CFisUC, Departamento de F\'isica, Universidade de Coimbra, 3004-516 Coimbra, Portugal
}
\date{The date of receipt and acceptance will be inserted by the editor}

\abstract{Co-orbital planets (in a 1:1 mean motion resonance) can be formed within a Laplace resonance chain. Here, we develop a secular model to study the dynamics of the resonance chain $p:p:p+1$, where the co-orbital pair is in a first-order mean motion resonance with the outermost third planet. Our model takes into account tidal dissipation through the use of a Hamiltonian version of the constant time-lag model, which extends the Hamiltonian formalism of the point-mass case. We show the existence of several families of equilibria, and how these equilibria extend to the complete system. In one family, which we call the \textit{main branch}, a secular resonance between the libration frequency of the co-orbitals and the precession frequency of the pericentres has unexpected dynamical consequences when tidal dissipation is added. We report the existence of two distinct mechanisms that make co-orbital planets much more stable within the $p:p:p+1$ resonance chain rather than outside it. The first one is due to negative real parts of the eigenvalues of the linearised system with tides, in the region of the secular resonance mentioned above. The second one comes from non-linear contributions of the vector field and it is due to eccentricity damping. These two stabilising mechanisms increase the chances of a still-to-come detection of exoplanets in the co-orbital configuration.}


\maketitle

\section{Introduction}
\label{S1}

Co-orbital systems have been extensively studied, since the discovery of equilibria in the three-body problem by \cite{Euler1764} and \cite{Lagrange1772}.  In the hierarchical case, that is, when $\left(m_1+m_2\right)/m_0<1/27$, where $m_0$ is the mass of the central body and $m_1$ and $m_2$ are the co-orbital masses, \cite{Ga1843} has shown that the equilateral equilibria, where the bodies are at the vertices of an equilateral triangle, are linearly stable. This result, combined with the discovery of several examples of co-orbital bodies, such as the Jovian trojans or the so-called horseshoe-shaped orbit of Janus \& Epimetheus around Saturn, contributed to increase the interest of scientists for this kind of systems.

In the planar and circular case, for masses in the range $3\times10^{-4}<\left(m_1+m_2\right)/m_0<1/27$, the angle $\lambda_1-\lambda_2$ between the co-orbitals librates around its equilibrium of $\pm60^{\circ}$ in orbits commonly called tadpole, where the co-orbital angle is bounded by\footnote{The exact value of the lower bound is $2\arcsin\left[\left(\sqrt{2}-1\right)/2\right]\approx23.9^{\circ}$ \citep[e.g.][]{RobutelPousse2013}.} $23.9^{\circ}<\lambda_1-\lambda_2<180^{\circ}$. However, for masses $\left(m_1+m_2\right)/m_0<3\times10^{-4}$ \citep[e.g.][]{LauCha2002}, a separatrix in the phase space delimits a region of stable trajectories of another kind, generally said horseshoe-shaped, where the critical angle $\lambda_1-\lambda_2$ librates around $180^{\circ}$ with at least $312.2^{\circ}$ of amplitude. For small eccentricities and libration amplitudes, still in the planar case, it has been shown, numerically by \cite{Guippone2010} and analytically by \cite{RobutelPousse2013}, the existence of two proper modes called Lagrange and anti-Lagrange. In the Lagrange configuration, the pericentres of the orbits do not precess and verify the relation $\varpi_1-\varpi_2=60^{\circ}$, whereas in the anti-Lagrange configuration, both orbits precess at the same frequency while maintaining the relation $\varpi_1-\varpi_2=240^{\circ}$. For small eccentricities but very large libration amplitudes, in the region of horseshoe-shaped orbits, \cite{CoRoCo2021} showed that the Lagrange and anti-Lagrange configurations correspond to $\varpi_1-\varpi_2=0^{\circ}$ and $\varpi_1-\varpi_2=180^{\circ}$, respectively. High eccentricities give rise to topological changes in the phase space \citep[see][]{LeRoCo2018} and thus to many more exotic trajectories, while the dynamics of the inclined problem is even more complex by allowing, among other things, transitions between the aforementioned orbits and retrograde co-orbitals \citep{Namouni1999}.

The discovery of exoplanets raised the question of the existence of co-orbital planets, which are absent from the solar system. Accretion in situ at the equilateral equilibria of a primary or capture in the 1:1 resonance of planets formed in other parts of the system are two possible scenarii of formation of such systems \citep[e.g.][]{LauCha2002, Cresswell_Nelson_2009}. The stability of co-orbital planets formed in a disk has been studied by \cite{Leleu2019}, who showed that under dissipative interactions with the gas disk, the equilateral equilibria can be either attractive or repulsive, depending on the co-orbital mass ratio and the parameters of the disk. Moreover, \cite{Leleu2019} showed that, at least around low-mass stars, co-orbital exoplanets generally end up in a tadpole configuration and often within a Laplace resonance chain.

For co-orbital exoplanets orbiting close to their host star, tidal dissipation induced by the differential gravitational interaction leads to a long term evolution of the orbits. \cite{CoRoCo2021} has shown that, for a pair of co-orbital exoplanets orbiting a star, the equilateral Lagrangian equilibria are always repulsive under tidal interactions, and that regardless of the parameters of the system, the destruction of the co-orbital motion is unavoidable. However, the discovery of co-orbital exoplanets is still possible because the destruction time is strongly dependent on the parameters of the system and can easily be larger than the lifetime of the host star. \cite{CoRoCo2021} neglected any interaction with possible other planets in the system. In this paper we extend this work to the case where the pair of co-orbital exoplanets interacts with an outermost third planet, in a first order mean motion resonance with the co-orbitals. More precisely, we study the Laplace resonance chain $p:p:p+1$, where $p$ is a small integer.

In Sect. \ref{S2}, we study the point-mass $p:p:p+1$ resonance chain, in the absence of tides. We show how rich and complex the dynamics of this chain is, and we conclude the section with the presentation of the stability map of the chain. In Sect. \ref{S3}, we include tidal dissipation in the model by an extension of the Hamiltonian formalism. We study the linearised system in the vicinity of the equilibria, and by computing the real parts of the eigenvalues, we show the existence of a zone linearly stable, around the $1\!:\!1$ secular resonance between the libration frequency of the co-orbital and the precession frequency of the pericentres. In Sect. \ref{S4}, we compare the analytical results with numerical simulations. They confirm the results of Sects. \ref{S2} and \ref{S3} and highlight the existence of a stabilisation mechanism of the co-orbitals due to eccentricity damping. We discuss our results in Sect. \ref{S5}. In Table \ref{notation} of Appendix \ref{append_notation}, we list the notations used throughout this paper. Appendix \ref{append_simu} completes Sect. \ref{S4} with more numerical simulations and a complete discussion on the influence of the mass of the third planet on the co-orbital dynamics.

\section{The $p:p:p+1$ resonance chain}
\label{S2}

\subsection{The Hamiltonian of the problem}\label{sec_hamiltonian}

In this section, we study an occurrence of the point-mass planar four-body problem. We construct the Hamiltonian associated with the resonance chain $p:p:p+1$, where a central body, the star of mass $m_0$, is orbited by two co-orbital planets of mass $m_1$ and $m_2$, and a third planet, of mass $m_3$, further away from the star and in a first-order mean motion resonance with the pair of co-orbital planets. Although we write all equations for a general value of the integer $p$, the figures are restricted to the case $p=1$, where the nominal period of the third planet is twice that of the co-orbitals. For all planets, we define the quantities $\beta_j=m_0m_j/\left(m_0+m_j\right)$ and $\mu_j=\mathcal{G}\left(m_0+m_j\right)$, where $\mathcal{G}$ is the gravitational constant. 

\subsubsection{The averaged Hamiltonian}\label{sec_expansion_a}

In order to define a canonical coordinate system related to the semimajor axis $a_j$, the eccentricity $e_j$, the mean longitude $\lambda_j$ and the longitude of the pericentre $\varpi_j$ of planet $j$, we first consider Poincar\'e heliocentric coordinates ($\tilde{\Lambda}_j,\lambda_j,\tilde{D}_j,-\varpi_j$) where
\begin{equation}\label{Lambda_def}
\tilde{\Lambda}_j=\beta_j\sqrt{\mu_ja_j}\place\text{and}\place \tilde{D}_j=\tilde{\Lambda}_j\left(1-\sqrt{1-e_j^2}\right).
\end{equation}
In these coordinates, the Hamiltonian derives from the symplectic form
\begin{equation}
\Omega=\sum_{j\in\left\lbrace 1,2,3\right\rbrace}\left(d\lambda_j\wedge d\tilde{\Lambda}_j-d\varpi_j\wedge d\tilde{D}_j\right).
\end{equation}
Following \cite{LaskarRobutel1995}, the planetary Hamiltonian is written
\begin{equation}\label{complete_hamiltonian}
H=H_K(\tilde{\Lambda}_j)+H_P(\tilde{\Lambda}_j,\lambda_j,\tilde{D}_j,\varpi_j),
\end{equation}
where the Keplerian part, due to star-planet interactions, reads
\begin{equation}
H_K=-\sum_{j\in\left\lbrace 1,2,3\right\rbrace}\frac{\beta_j^3\mu_j^2}{2\tilde{\Lambda}_j^2},
\end{equation}
and the perturbation $H_P$, whose size relative to $H_K$ is of order $\left(m_1+m_2+m_3\right)/m_0$, is expanded in power series of the eccentricities. We assume that the system is close to the resonance $p:p:p+1$. This means that the nominal mean motions verify
\begin{equation}\label{nominal}
n_{1,0}=n_{2,0}=\eta=\frac{p+1}{p}n_{3,0},
\end{equation}
while the nominal semimajor axes $a_{j,0}$ are related to $n_{j,0}$  by the ``Kepler law'' $n_{j,0}^2a_{j,0}^3 = \mu_0 = \mathcal{G}m_0$. The $a_j$ are always close to their nominal value\footnote{The nominal semimajor axes are defined with $\mu_0$ instead of $\mu_j$, which conveniently yields $a_{1,0}=a_{2,0}$. This approximation is valid since the subsequent error is of order $\mathcal{O}\left(m_j/m_0\right)$, while the width of the resonance is of order $\mathcal{O}\sqrt{m_j/m_0}$.}
$a_{j,0}$, and the $\tilde{\Lambda}_j$ stay close to the quantities $\Lambda_j^{\star}$ defined as
\begin{equation}\label{resonance_keplerienne}
\Lambda_j^{\star}=m_j\sqrt{\mu_0a_{j,0}}.
\end{equation}
To study the dynamics in the vicinity of the resonance, we expand the Hamiltonian in the neighbourhood of $\left(\Lambda_1^{\star},\Lambda_2^{\star},\Lambda_3^{\star}\right)$. An expansion at order $2$ in the Keplerian part and at order $0$ in the perturbative part generates remainders of the same size and we limit ourselves to
\begin{equation}
H_P(\tilde{\Lambda}_j,\lambda_j,\tilde{D}_j,\varpi_j)=H_P(\Lambda_j^{\star},\lambda_j,\tilde{D}_j,\varpi_j).
\end{equation}
A suitable linear change of variables to deal with the $p:p:p+1$ resonance chain is \citep[e.g.][]{Delisle2017}
\begin{equation}\label{change_var_angle}
\begin{pmatrix}\xi\\\xi_2\\\xi_3\\\sigma_1\\\sigma_2\\\sigma_3\end{pmatrix}=\begin{pmatrix}
1&-1&0&0&0&0\\
0&p&-p&0&0&0\\
0&-p&p+1&0&0&0\\
0&-p&p+1&1&0&0\\
0&-p&p+1&0&1&0\\
0&-p&p+1&0&0&1
\end{pmatrix}
\begin{pmatrix}
\lambda_1\\\lambda_2\\\lambda_3\\-\varpi_1\\-\varpi_2\\-\varpi_3
\end{pmatrix},
\end{equation}
which is canonical if we transform the actions according to
\begin{equation}\label{change_var_action}
\begin{split}
&\left(\tilde{\Lambda}_1,\tilde{\Lambda}_2,\tilde{\Lambda}_3,\tilde{D}_j\right)\mapsto\left(L',\Gamma',G',D_j'\right)=\\
&\left(\tilde{\Lambda}_1,\frac{p+1}{p}\left(\tilde{\Lambda}_1+\tilde{\Lambda}_2\right)+\tilde{\Lambda}_3,\sum_{j\leq 3}\left(\tilde{\Lambda}_j-\tilde{D}_j\right),\tilde{D}_j\right).
\end{split}
\end{equation}
Since the total angular momentum $G'$ is a first integral, the Hamiltonian does not depend on the angle $\xi_3$. Moreover, in the $p:p:p+1$ resonance, the angle $\xi_2$ is fast circulating and we average over it. The averaged Hamiltonian reads
\begin{equation}\label{average}
\begin{split}
&H'\left(L',\Gamma',G',\xi,D_j',\sigma_j\right)=\\&H_K'\left(L',\Gamma',G',D_j'\right)+\frac{1}{2\pi}\int_0^{2\pi}H_P'\left(\xi,\xi_2,D_j',\sigma_j\right)d\xi_2.
\end{split}
\end{equation}
This change of variable, along with the averaging process, allows to lose the $2$ degrees of freedom associated with $\left(\xi_2,\xi_3,\Gamma',G'\right)$ and we are left with $4$ degrees of freedom.
After the averaging process, the scaling factor $\Gamma'$ and the angular momentum $G'$ are both parameters and a rescaling by $\Gamma'$ reduces the dependency to only one parameter. As we study the effect of tidal dissipation on the dynamics in Sect. \ref{S3}, it is actually more convenient to normalise by the constant $\Gamma^{\star}=(p+1)\left(\Lambda_1^{\star}+\Lambda_2^{\star}\right)/p+\Lambda_3^{\star}$, rather than by $\Gamma'$, which is not constant when dissipation is present. That is, we perform the canonical transformation
\begin{equation}\label{normalization}
\begin{split}
&\Ham=\frac{H'}{\Gamma^{\star}},\;\;L=\frac{L'}{\Gamma^{\star}},\;\;G=\frac{G'}{\Gamma^{\star}},\\&\Gamma=\frac{\Gamma'}{\Gamma^{\star}},\;\;D_j=\frac{D_j'}{\Gamma^{\star}},\;\;\Lambda_j=\frac{\tilde{\Lambda}_j}{\Gamma^{\star}},
\end{split}
\end{equation}
while the angles are unchanged.

\subsubsection{Expansion of the Keplerian part}
As mentioned above, the Keplerian part of the Hamiltonian is expanded at second order in the vicinity of the $\Lambda_j^{\star}$. If we note $\Delta\tilde{\Lambda}_j=\tilde{\Lambda}_j-\Lambda_j^{\star}$, the expansion reads
\begin{equation}\label{HK_petit_sans_change_var}
H_K=\sum_{j=1}^3 n_{j,0}\Delta\tilde{\Lambda}_j-\frac{3}{2}\sum_{j=1}^3 \frac{n_{j,0}}{\Lambda_j^{\star}}\Delta\tilde{\Lambda}_j^2.
\end{equation}
Substituting $\tilde{\Lambda}_j-\Lambda_j^{\star}$ for $\Delta\tilde{\Lambda}_j$, we obtain
\begin{equation}
H_K=4\sum_{j=1}^3n_{j,0}\tilde{\Lambda}_j-\frac{3}{2}\sum_{j=1}^3\frac{n_{j,0}}{\Lambda_j^{\star}}\tilde{\Lambda}_j^2-\frac{5}{2}\sum_{j=1}^3n_{j,0}\Lambda_j^{\star}.
\end{equation}
The third term is constant and can be removed without changing the dynamics. Performing the normalisation (\ref{normalization}) and the change of variable (\ref{change_var_action}), we have
\begin{equation}\label{HK_entier}
\begin{split}
&\Ham_K=-\frac{3}{2}\eta\left\lbrace\accolade C_1L^2+C_2\left( p\left(\Gamma-\Upsilon\right)-L\right)^2\right.\\
& \left. \place\place+C_3p\left(p+1\right)\left(\Upsilon-\frac{p\Gamma}{p+1}\right)^2\right\rbrace+\frac{4\eta p\Gamma}{p+1},
\end{split}
\end{equation}
where we noted $\Upsilon=G+D_1+D_2+D_3=\sum_j\Lambda_j$ and $C_j=\Gamma^{\star}/\Lambda_j^{\star}$, that is
\begin{equation}\label{C_j}
\begin{split}
&C_1=\left(\frac{p+1}{p}\right)^{1/3}\frac{m_3}{m_1}+\frac{p+1}{p}\left(1+\frac{m_2}{m_1}\right),\\
&C_2=\frac{m_1}{m_2}C_1,\;\;C_3=1+\left(\frac{p+1}{p}\right)^{2/3}\frac{m_1+m_2}{m_3}.
\end{split}
\end{equation}
Without dissipation, $\Gamma$ is constant and almost equal to $1$ and we simply evaluate $\Ham_K$ at $\Gamma=1$, hence achieving the reduction to only one parameter\footnote{The relevant parameter to consider is $\Gamma/G$ and the variations of $\Gamma$ are reported in $G$ (Eqs. (\ref{delta_def}) and (\ref{Deltag_def})).}. In this case, the last term is constant and can also be removed.

Instead of the variables $L$, $G$ and $\Gamma$, we can use the variables $\Delta L$, $\Delta G$ and $\Delta \Gamma$ defined by their difference to the Keplerian resonance (\ref{resonance_keplerienne}). In that case, the approximation $\Gamma=1$ becomes $\Delta \Gamma=0$ and Eq. (\ref{HK_petit_sans_change_var}) yields, once normalised
\begin{equation}\label{HK_petit_avec_change_var}
\begin{split}
&\Ham_K=-\frac{3}{2}\eta\left\lbrace\left(p^2C_2+p(p+1)C_3\right)\Delta\Upsilon^2\right.\\
&\left. \place\place+2pC_2\Delta\Upsilon\Delta L+\left(C_1+C_2\right)\Delta L^2\right\rbrace,
\end{split}
\end{equation} 
where $\Delta\Upsilon=\Delta G+D_1+D_2+D_3=\sum_j\Delta\Lambda_j$. We find the Hamiltonian (\ref{HK_petit_avec_change_var}) to be well adapted to the analytical work derived in Sect. \ref{sec_analytical}, while we rather use the Hamiltonian (\ref{HK_entier}) in the remaining sections. Moreover, we do not perform the evaluation $\Gamma=1$ in Sect. \ref{S3}, where tidal dissipation is present and $\Gamma$ is a variable quantity. Both Hamiltonians yield the same dynamics and it is only a matter of preference.

\subsubsection{Expansion of the perturbative part}\label{sec_expansion_e}The perturbative part $\Ham_P$ of the Hamiltonian is expanded in power series of the eccentricities. To this aim, we separate the contributions due to interactions between each pair of planets as \begin{equation}
\Ham_P=\Ham_{1,2}+\Ham_{1,3}+\Ham_{2,3}.
\end{equation}
We note $X_j=\sqrt{2\tilde{D}_j/\tilde{\Lambda}_j}\,e^{i\varpi_j}=e_je^{i\varpi}+\gO{e_j^3}$. For a couple $\left(p_1,p_2\right)\in\left\lbrace\left(1,2\right),\left(1,3\right),\left(2,3\right)\right\rbrace$ of planets, the perturbation to the Hamiltonian due to their mutual interaction reads \citep{LaskarRobutel1995}
\begin{equation}
\Ham_{p_1,p_2}\!\!=\!\!\!\sum_{\vect{k}\in\mathbb{Z}^2}\!\!\left(\!\sum_{\vect{q}\in\mathbb{N}^4}\!\!\Psi_{\vect{k},\vect{q}}X_{p_1}^{q_1}X_{p_2}^{q_2}\Xb_{p_1}^{q_3}\Xb_{p_2}^{q_4}\!\!\right)\!\!e^{i\left(k_1\lambda_{p_1}+k_2\lambda_{p_2}\right)}.
\end{equation}
For a non-zero $\Psi_{\vect{k},\vect{q}}$, the conservation of the angular momentum imposes on the tuples $\vect{q}=\left(q_1,q_2,q_3,q_4\right)\in\mathbb{N}^4$ and $\vect{k}=\left(k_1,k_2\right)\in\mathbb{Z}^2$ to verify the so-called d'Alembert rule
\begin{equation}
k_1+k_2+q_1+q_2-q_3-q_4=0.
\end{equation}
This rule, combined with the averaging process, implies that $\Ham_{1,2}$ has no odd term in eccentricity, while $\Ham_{1,3}$ and $\Ham_{2,3}$ have no term of order $0$. Since we limit ourselves to the second order in eccentricity, we write
\begin{equation}
\Ham_{1,2}=\Ham^{(0)}+\Ham_{1,2}^{(2)},\place \Ham_{j,3}=\Ham_{j,3}^{(1)}+\Ham_{j,3}^{(2)},
\end{equation}
where the superscript refers to the order in eccentricity while the subscript refers to the considered couple of planets. $\Ham^{(0)}$ has no subscript since only the pair of co-orbitals yields terms of order $0$ and no confusion is possible.

Following \cite{LaskarRobutel1995}, $\Ham_{1,3}$ and $\Ham_{2,3}$ can be written
%
\begin{equation}\label{O1_j3}
\begin{split}
&\Ham_{j,3}^{(1)}=\frac{m_jn_{3,0}}{m_0C_3}\left\lbrace C_{p,1}^{(1)}\sqrt{2C_jD_j}\cos\left(p\delta_{j,1}\xi -\sigma_j\right)\right.\!\!\!\!\!\!\!\!\!\!\!\!\\ 
&\left. \place\place\;+C_{p,2}^{(1)}\sqrt{2C_3D_3}\cos\left(p\delta_{j,1}\xi -\sigma_3\right)\right\rbrace\\
\end{split}
\end{equation}
and
%
\begin{equation}\label{O2_j3}
\begin{split}
&\Ham_{j,3}^{(2)}=\frac{2m_j}{m_0}\frac{n_{3,0}}{C_3}\left\lbrace C_{p,1}^{(2)}C_jD_j\cos\left(2p\delta_{j,1}\xi -2\sigma_j\right)\right.\\
& +C_{p,2}^{(2)}C_3D_3\cos\left(2p\delta_{j,1}\xi -2\sigma_3\right)\\
& +C_{p,3}^{(2)}\sqrt{C_jC_3D_jD_3}\cos\left(2p\delta_{j,1}\xi - \sigma_j - \sigma_3\right)\!\!\!\!\!\!\!\!\!\!\\
&+C_{p,4}^{(2)}\left(C_jD_j+C_3D_3\right)\\
&\left. +C_{p,5}^{(2)}\sqrt{C_jD_jC_3D_3}\cos\left(\sigma_j - \sigma_3\right)\right\rbrace,
\end{split}
\end{equation}
where $\delta_{j,1}=1$ if $j=1$ and zero otherwise.
The quantities $C_{p,m}^{(n)}$
depend only on $p$ and can be obtained using the Laplace coefficients. For $p=1$, their analytical expressions, as well as a numerical evaluation, is given in appendix \ref{append_coefficient}.

The perturbation $\Ham_{1,2}$ cannot be obtained using the same procedure, since  the Laplace coefficients diverge in $1$ and the two co-orbitals have the same nominal semimajor axes. We rather follow the method described in \cite{RobutelPousse2013}. We note $\Delta=\sqrt{2-2\cos\xi}$ and find
\begin{equation}
\begin{split}
&\Ham_0=\frac{m_1}{m_0}\frac{\eta}{C_2}\left(\cos\xi-\Delta^{-1}\right),\\
&\Ham_{1,2}^{(2)}=\frac{m_1}{m_0}\frac{\eta}{C_2}\left\lbrace A_h\left(C_1D_1+C_2D_2\right)\right.\\
&\left. \place\place+2\sqrt{C_1C_2D_1D_2}\,\mathcal{R}_{\text{e}}\left(B_he^{i\left(\sigma_2-\sigma_1\right)}\right)\right\rbrace,
\end{split}
\end{equation}
where
\begin{equation}
\begin{split}
&A_h=\frac{5\cos 2\xi-13+8 \cos \xi}{4\Delta^5}-\cos\xi\;\;\text{ and}\\&B_h=e^{-2i\xi}-\frac{ e^{-3i\xi}+16e^{-2i\xi}-26e^{-i\xi}+9e^{i\xi} }{8\Delta^5}.
\end{split}
\end{equation}
The final simplified Hamiltonian is then
\begin{equation}\label{total_hamiltonian}
\Ham=\Ham_K+\Ham^{(0)}+\Ham_{1,3}^{(1)}+\Ham_{2,3}^{(1)}+\Ham_{1,2}^{(2)}+\Ham_{1,3}^{(2)}+\Ham_{2,3}^{(2)}.
\end{equation}
We note $F_0:\mathbb{R}^8\mapsto\mathbb{R}^8$ the differential system derived from Eq. (\ref{total_hamiltonian}) by the Hamilton-Jacobi equations.


\subsection{Equilibria and linearisation in their vicinity}\label{sec_equilibria}

In this section, we study the equilibria of the resonance $p:p:p+1$ and the dynamics in their vicinity.

\subsubsection{Fixed points and libration centres}\label{pf_vs_centre_libration}

One of the consequences of averaging over the mean motion is that the averaged Hamiltonian (\ref{total_hamiltonian}) has equilibria, that is, points in the phase space where its gradient vanishes. The complete Hamiltonian (\ref{complete_hamiltonian}) though, has no equilibria, and since the Hamiltonian (\ref{total_hamiltonian}) is supposed to model it, we are interested in the dynamics of the complete Hamiltonian at the equilibria of the model.

At a fixed point (or equilibrium) of the model, $L$ and the $D_j$ are constant and so are the $e_j$ and the $a_j$. Similarly, the angles $\sigma_j$ and $\xi$ are constant, that is, there exists constants $c_j$ such that\begin{equation}
\sigma_j=-p\lambda_2+\left(p+1\right)\lambda_3-\varpi_j=c_j.
\end{equation} 
However, the secular angle $-p\lambda_2+\left(p+1\right)\lambda_3$ and the pericentres $\varpi_j$ are not constant at the equilibria, but they all precess with the same frequency which we note $\nu_3$.

The average performed in (\ref{average}) is actually analogue to a first-order Lie serie expansion and the averaging process can be seen as a periodic change of variable. Indeed, noting $x$ and $x'$ the variables of the Hamiltonian respectively before and after the average, one has $x=e^{L_W}x'$ where $L_W=\left\lbrace W,\cdot\right\rbrace$ denotes the total time derivative along the trajectories of the scalar field $W$, which is constrained by the cohomological equation \citep{Deprit1969}
\begin{equation}
\left\lbrace H_K,W\right\rbrace=H_P-\frac{1}{2\pi}\int_0^{2\pi}H_P\,d\xi_2.
\end{equation}
This equation shows that, at the equilibria, $W$ is periodic of time, and so is the change of variable. That is, fixed points in the averaged model correspond to periodic trajectories in the complete system. For a quantity not invariant by rotation around the axis of the total angular momentum, though, like the secular angle $-p\lambda_2+\left(p+1\right)\lambda_3$ or the pericentres $\varpi_j$, a fixed point in the model corresponds in the complete Hamiltonian to a quasi-periodic motion with the two frequencies $\nu_2$ and $\nu_3$, with
\begin{equation}\label{nu_23}
\nu_2=\dot{\xi}_2=\drond{\Ham}{\Gamma}\place\text{and}\place \nu_3=\dot{\xi}_3=\drond{\Ham}{G}=\drond{\Ham}{\Upsilon},
\end{equation}
where these quantities are evaluated at the equilibrium. More precisely, it is a periodic motion with frequency $\nu_2$ in a rotating frame following all the pericentres at frequency $\nu_3$. This result holds true for any resonance chain \citep[e.g. Eq. (A.1) of][]{Delisle2017}. In the rest of this work, what is referred to as a fixed point, or equilibrium, for the model, will be referred to as a \textit{libration centre} in the complete system.

\subsubsection{Analytical results at first order in eccentricity}\label{sec_analytical}

Even if truncated at order $1$ in eccentricity, the fixed points of the Hamiltonian (\ref{total_hamiltonian}) cannot be given analytically. Similar difficulties were met by \cite{Delisle2017} for resonance chains with first order resonances between non-consecutive planets. However, we show here that a further simplification allows to obtain analytical expressions of the equilibria and of the eigenvalues of the linearised system.
\begin{figure*}[h]
	\centering
	\includegraphics[width=1\linewidth]{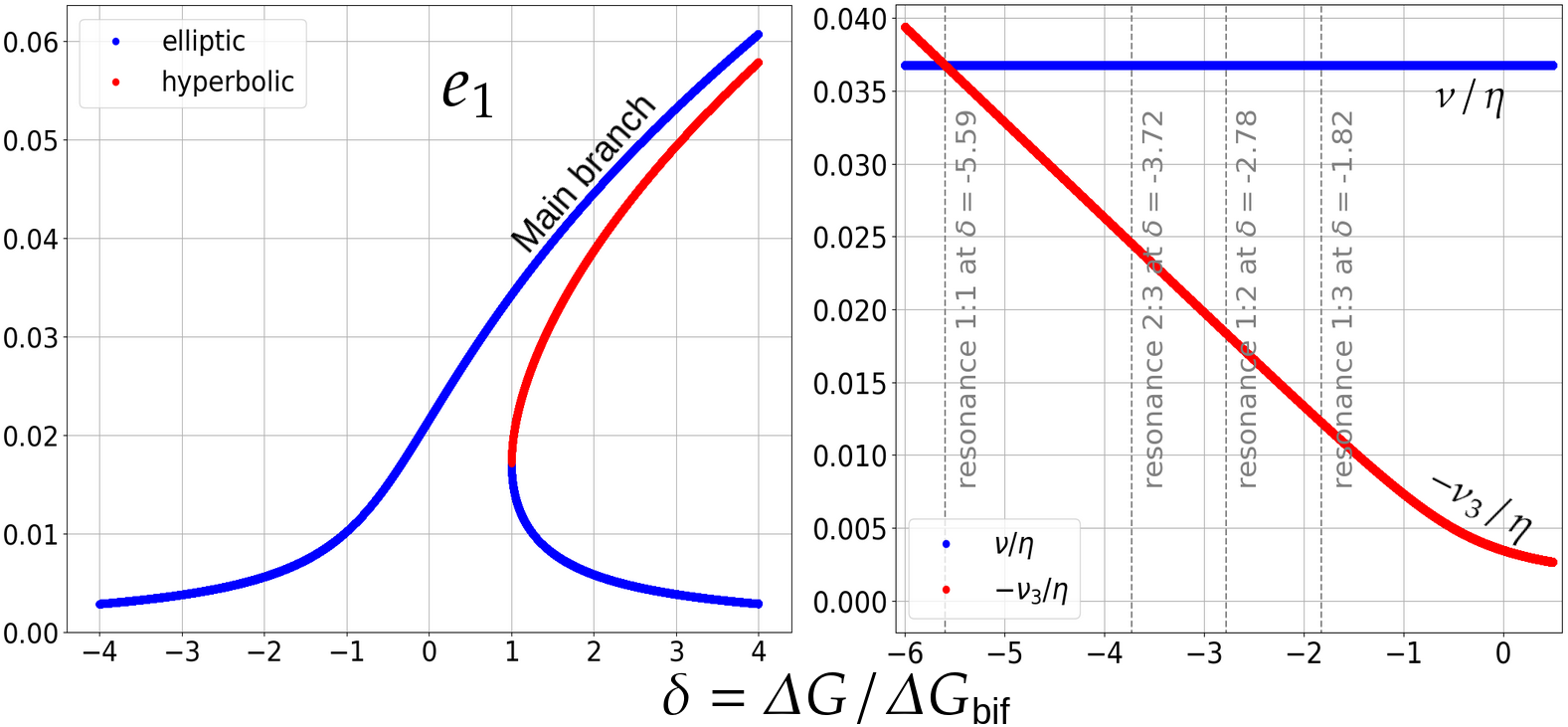}
	\caption{Left : Value of $e_1$ at the fixed points of the resonance chain $1\!:\!1\!:\!2$ as predicted by Eq. (\ref{poly_deg_3}). Right : Values of $\nu/\eta$ and $\nu_3/\eta$ along the main branch as a function of $\delta$, for the same resonance chain, predicted by Eqs. (\ref{nu}) and (\ref{nu3}). In both panels the planetary masses are $\left(m_1+m_2\right)/2=m_3=10^{-4}\,m_0$ and $m_2/m_1=10$. According to Eq. (\ref{pf_O1_action}), we have $e_1=e_2$, and, for this choice of masses and resonance chain, $e_3/e_1=0.5471$. In Table \ref{tableau_de_pf}, which gathers the fixed points at second order in eccentricity, the three branches visible on the left plot are the branches $1$ (main branch), $2$ and $6$. The secular resonances between $\nu$ and $\nu_3$ are shown on the right, and are also easy to spot on Fig. \ref{fig_map}.}\label{fig_e1_O1+reso_nu_nu3}
\end{figure*}

To further simplify the Hamiltonian, we force a decoupling between the degree of freedom $\left(\xi,\Delta L\right)$ associated with the libration of the co-orbitals and the three other degrees of freedom $\left(\sigma_j,D_j\right)$. To this end, we first evaluate $\Ham^{(1)}=\Ham_{1,3}^{(1)}+\Ham_{2,3}^{(1)}$ at $\xi=\pi/3$. Indeed, the Hamiltonian $\Ham_K+\Ham^{(0)}+\Ham^{(1)}$ is only a perturbation of $\Ham_K+\Ham^{(0)}$ which only has equilibria at $\xi\in\left\lbrace\pm\pi/3,\pi\right\rbrace$, where $\xi=\pi$ is the hyperbolic (unstable) aligned configuration and $\xi=\pm\pi/3$ are the equilateral elliptic (stable) equilibria \citep{RobutelPousse2013}. Both elliptic equilibria are symmetric with the same dynamics, hence we only consider $\xi=\pi/3$. Then, we replace the variable $\Delta L$ by the constant $\Delta L^{\star}$ in the anti-diagonal term\footnote{That is, the term proportional to $\Delta L\Delta \Upsilon$} of $\Ham_K$ in (\ref{HK_petit_avec_change_var}), where
\begin{equation}\label{decouplage}
\begin{split}
&\Delta L^{\star}=-\frac{pC_2}{C_1+C_2}\left(\Delta G+D_{1,0}+D_{2,0}+D_{3,0}\right)\\&\place\;\;=-\frac{pC_2}{C_1+C_2}\Delta\Upsilon^{\star},\end{split}
\end{equation}
is the value of $\Delta L$ for which $\partial\Ham_K/\partial\Delta L$ vanishes. The $D_{j,0}$ are given by Eq. (\ref{pf_O1_action}). While the evaluation at $\xi=\pi/3$ allows analytical expressions for the position of the fixed points, the evaluation at $\Delta L=\Delta L^{\star}$ also uncouples $\left(\xi,\Delta L\right)$ from $\left(\sigma_j,D_j\right)$ and enables analytical expressions of the eigenvalues of the linearised system in the vicinity of the fixed points.
\begin{table*}
	\begin{center}
		\begin{tabular}{rrrrrrrrrrr}
			\hline
			\#&$100\,e_1$&$100\,e_2$&$100\,e_3$&$\sigma_1$ ($^{\circ}$)&$\sigma_2$ ($^{\circ}$)&$\sigma_3$ ($^{\circ}$)&{\small$L\!-\!0.0345$}&$\xi$ ($^{\circ}$)&nature&domain\\
			\hline
			1&$4.449$&$7.490$&$5.878$&$\pm 14.433$&$\pm22.572$&$\mp92.014$&$6.461\text{e}\!-\!5$&$\pm59.760$&elliptic&$\delta\in\mathbb{R}$\\
			2&$0.168$&$0.165$&$0.093$&$\mp119.18$&$\pm179.92$&$\pm4.6606$&$19.34\text{e}\!-\!5$&$\pm60.003$&$\delta$-dependant&$\delta>1.129$\\
			3&$5.201$&$7.645$&$5.470$&$\pm14.216$&$\mp15.688$&$\pm95.525$&$6.483\text{e}\!-\!5$&$\pm59.806$&elliptic&$\delta>5.997$\\
			4&$8.344$&$8.646$&$0.305$&$180$&$0$&$0$&$6.501\text{e}\!-\!5$&$180$&hyperbolic&$\delta\in\mathbb{R}$\\
			5&$7.939$&$7.180$&$5.922$&$\mp151.76$&$\mp20.944$&$\pm92.729$&$6.439\text{e}\!-\!5$&$\pm179.13$&hyperbolic&$\delta>4.195$\\
			6&$10.11$&$6.330$&$6.246$&$\mp102.00$&$\pm177.92$&$\pm4.3287$&$7.200\text{e}\!-\!5$&$\pm62.538$&hyperbolic&$\delta>1.129$\\
			7&$6.009$&$8.721$&$1.696$&$\pm6.8311$&$\mp1.8178$&$\pm86.449$&$6.518\text{e}\!-\!5$&$\pm59.710$&hyperbolic&$\delta>5.999$\\
			8&$0.164$&$0.165$&$0.079$&$0$&$180$&$0$&$19.34\text{e}\!-\!5$&$180$&hyperbolic&$\delta>1.082$\\
			9&$3.708$&$6.584$&$6.887$&$0$&$180$&$0$&$7.157\text{e}\!-\!5$&$180$&hyperbolic&$\delta>1.082$\\
			\hline
		\end{tabular}
		\caption{The $15$ equilibria of the simplified Hamiltonian (\ref{total_hamiltonian}).}\label{tableau_de_pf}
	\end{center}
	{\small The equilibria are found for the resonance chain $1\!:\!1\!:\!2$ at $\delta=7$ using a Newton-Raphson method. Values given without decimal places are exact. The planetary masses are as in Fig. \ref{fig_e1_O1+reso_nu_nu3}. Branch $2$ is hyperbolic only for $5.548\leq\delta\leq5.802$ and elliptic elsewhere. The entry value of $\delta$ in the formal resonance (here $1.129$) weakly depends on the planetary masses, because of the normalisation by $\Delta G_{\text{bif}}$. Branch $1$ is the only elliptic branch existing for all values of $\delta$ and it is the main branch introduced in Sect. \ref{sec_analytical}. It corresponds to the only real solution of Eq. (\ref{poly_deg_3}) when $\delta<1$. Branches $3$, $5$ and $7$ do not exist at first order in eccentricity, while they exist at second order, hence, we cannot exclude that the complete Hamiltonian (\ref{complete_hamiltonian}) has more \textit{libration centres}, either because we did not discretise the phase space thinly enough to find them, or because they do not exist at second order in eccentricity.}
\end{table*}
The differential system derived from $\Ham_K+\Ham^{(0)}+\Ham^{(1)}$, once these simplifications have been performed, is given in appendix \ref{append_syst_diff}. It vanishes when the angles are equal to
\begin{equation}\label{pf_O1_angle}
\begin{split}
&\xi_0=\frac{\pi}{3},\place\sigma_{1,0}=p\frac{\pi}{3}+\epsilon\pi,\place \sigma_{2,0}=\epsilon\pi,\\&\sigma_{3,0}=\arctan\frac{m_1\sin p\xi_0}{m_2+m_1\cos p\xi_0}+\left(1-\epsilon\right)\pi,
\end{split}
\end{equation}
where
\begin{equation}
\epsilon=\begin{cases}
0 & \text{if }\nu_3<0,\\
1 & \text{if }\nu_3>0,
\end{cases}
\end{equation}
and when the actions are equal to\footnote{$C_1D_{1,0}=C_2D_{2,0}$ yields $e_{1,0}=e_{2,0}$, since $e_j=\sqrt{2C_jD_j}$.}
\begin{equation}\label{pf_O1_action}
\begin{split}
&\frac{C_1{C_{p,1}^{(1)}}^2m_1^2p^2}{2C_3^2m_0^2\left(p+1\right)^2}=\left(\frac{\nu_3}{\eta}\right)^2D_{1,0},\;\; C_1D_{1,0}=C_2D_{2,0},\\&\frac{D_{3,0}}{D_{1,0}}=\frac{C_3{C_{p,2}^{(1)}}^2H^2}{C_1{C_{p,1}^{(1)}}^2},\;\; \Delta L_0=0,
\end{split}
\end{equation}
where the precession frequency of the pericentres is
\begin{equation}\label{nu3}
\nu_3=-\eta K\Delta\Upsilon^{\star},\place K=\frac{3p^2C_1C_2}{C_1+C_2}+3p\left(p+1\right)C_3,
\end{equation}
and we defined the constant $H$ by
\begin{equation}
H=\cos\left(p\xi_0-\sigma_{3,0}\right)+\frac{m_2}{m_1}\cos\sigma_{3,0}.\end{equation}
The unknowns of Eq. (\ref{pf_O1_action}) are the $D_{j,0}$, and since the ratios $D_{2,0}/D_{1,0}$ and $D_{3,0}/D_{1,0}$ are known, we are reduced to the unique unknown $D_{1,0}$. The precession frequency of the pericentre, $\nu_3$, depends on $D_{1,0}$ (see Eqs. (\ref{decouplage}) and (\ref{nu3})). Denoting $C=1+D_{2,0}/D_{1,0}+D_{3,0}/D_{1,0}$ and performing the translation $Z=D_{1,0}+2\Delta G/3C$, Eq. (\ref{pf_O1_action}) is rewritten as a third degree polynomial in $Z$
\begin{equation}\label{poly_deg_3}
\begin{split}
&Z^3-PZ-Q=0,\;\;\text{where}\\&P=\frac{\Delta G^2}{3C^2}\text{ and }Q=\frac{2\Delta G^3}{27C^3}+\frac{C_1{C_{p,1}^{(1)}}^2m_1^2p^2}{2C^2C_3^2K^2m_0^2\left(p+1\right)^2}.
\end{split}
\end{equation}
The coefficients $P$ and $Q$ of this polynomial depend on the parameter $\Delta G$. There is a bifurcation between $1$ and $3$ real solutions when $27Q^2-4P^3=0$, that is at
\begin{equation}\label{Dgbif}
\Delta G=\Delta G_{\text{bif}}=-\left(\frac{27C_1{C_{p,1}^{(1)}}^2p^2Cm_1^2}{8C_3^2K^2m_0^2\left(p+1\right)^2}\right)^{1/3}.
\end{equation}
In the rest of this work, we use the parameter
\begin{equation}\label{delta_def}
\delta=\Delta G/\Delta G_{\text{bif}},
\end{equation}
where
\begin{equation}\label{Deltag_def}
\Delta G=\frac{G'}{\Gamma'}-\frac{G^{\star}}{\Gamma^{\star}}=\frac{G}{\Gamma}-\frac{\sum\Lambda_j^{\star}}{\Gamma^{\star}}=\frac{G}{\Gamma}-\sum_{j\leq 3}C_j^{-1}.
\end{equation}
In this section, $\Gamma=\Gamma'/\Gamma^{\star}\approx1$ is simply evaluated at $1$ and ignored, but not in Sects. \ref{S3} and \ref{S4}, where tidal dissipation induces a drift in $\Gamma$, hence in $\delta$. The normalisation by $\Delta G_{\text{bif}}$ ensures that the bifurcation is at $\delta=1$, regardless of the planetary masses.

The forced decoupling that we performed to obtain these expressions allows us to end up with results very similar to the second fundamental model of resonance proposed by \cite{Henrard1983}. Indeed, the fixed points are given by the roots of the third degree polynomial in $Z$ (\ref{poly_deg_3}), which has $1$ or $3$ real solutions depending on $\delta$, hence a bifurcation. The solutions of Eq. (\ref{poly_deg_3}) are plotted in Fig. \ref{fig_e1_O1+reso_nu_nu3}. For $\delta<1$, only one elliptic equilibrium exists, called the main branch, while for $\delta\geq1$, two other fixed points appear, one of them being hyperbolic, hence the presence of separatrices in the phase space and the formal existence of a resonance. These results come from strong hypothesis and we see in Sect. \ref{sec_topology} that the topology of the Hamiltonian (\ref{total_hamiltonian}) is different (see Table \ref{tableau_de_pf}). However, we show in Fig. \ref{fig_O1_O2} that these analytical expressions are accurate for small eccentricities.

In the vicinity of the main branch, we linearise the differential system. We use the cartesian coordinates
\begin{equation}
u_j=\sqrt{2D_j}\cos\sigma_j\place\text{and}\place v_j=\sqrt{2D_j}\sin\sigma_j,
\end{equation}
and noting $X=\,^t\left(u_1,u_2,u_3,v_1,v_2,v_3,\Delta L,\xi\right)$, the linearised system is
\begin{equation}\label{linear_O1}
\begin{split}
&\frac{d\Delta X}{dt}=\begin{pmatrix}
\mathcal{Q}_6&0_{6,2}\\0_{2,6}&\mathcal{Q}_2
\end{pmatrix}\Delta X,\;\text{ where }\; \Delta X=X-X_0,\\&\mathcal{Q}_2=\begin{pmatrix}
0&\frac{9m_1}{4m_0}\eta C_2^{-1}\\-3\eta\left(C_1+C_2\right)&0
\end{pmatrix},
\end{split}
\end{equation}
and $X_0$ is the equilibrium value of $X$. The matrix $\mathcal{Q}_6$ is given in appendix \ref{Q6}. Its characteristic polynomial reads
\begin{equation}\label{chi}
\det\left(\lambda I_6\!-\!\mathcal{Q}_6\right)\!=\!\left(\lambda^2\!+\!\nu_3^2\right)^2\!\left(\lambda^2\!+\!\nu_3^2-2\nu_3I\sum D_{j,0}\right),
\end{equation}
where $I$ is defined in Eq. (\ref{notation_Q6}). It is interesting to note that the precession frequency of the pericentres, $\pm i\nu_3$, is an eigenvalue of the differential system (\ref{linear_O1}). 
This factorisation was already noticed by \cite{Pucacco2021}, who studied the resonance chain $1:2:4$ of the Galilean satellites, although it was not attributed to the precession of the pericentres. The eigenvalues of $\mathcal{Q}_2$ are $\pm i\nu$, where
\begin{equation}\label{nu}
\nu=\eta\sqrt{\frac{27}{4}\frac{m_1+m_2}{m_0}}
\end{equation}
is the libration frequency of the angle $\xi$ in the neighbourhood of the equilateral Lagrangian configuration \citep{RobutelPousse2013,CoRoCo2021}. Figs. \ref{fig_e1_O1+reso_nu_nu3} and \ref{fig_nu3_sur_eta} show that $\nu_3<0$ for the main branch, which ensures that the roots of (\ref{chi}) are pure imaginary. Evaluating the eigenvalues $\pm i\nu$ and $\pm i\nu_3$ in the vicinity of the main branch shows that at $\delta\approx-5.6$ for the planetary masses in Fig. \ref{fig_e1_O1+reso_nu_nu3}, all these eigenvalues have roughly the same value, yielding a $1\!:\!1$ secular resonance between the libration frequency of the co-orbitals and the precession frequency of the pericentres. Other secular resonances between $\nu$ and $\nu_3$ are shown in Fig. \ref{fig_e1_O1+reso_nu_nu3}, and are also very visible on the stability map from Fig.~\ref{fig_map}. We show in Sect. \ref{sec_pseudo_fixed} that the secular resonance $1:1$ has important consequences for the tidal stability of the co-orbital pair.

\subsubsection{Topology of the phase space}\label{sec_topology}
\begin{figure*}[h]
	\centering
	\includegraphics[width=1\linewidth]{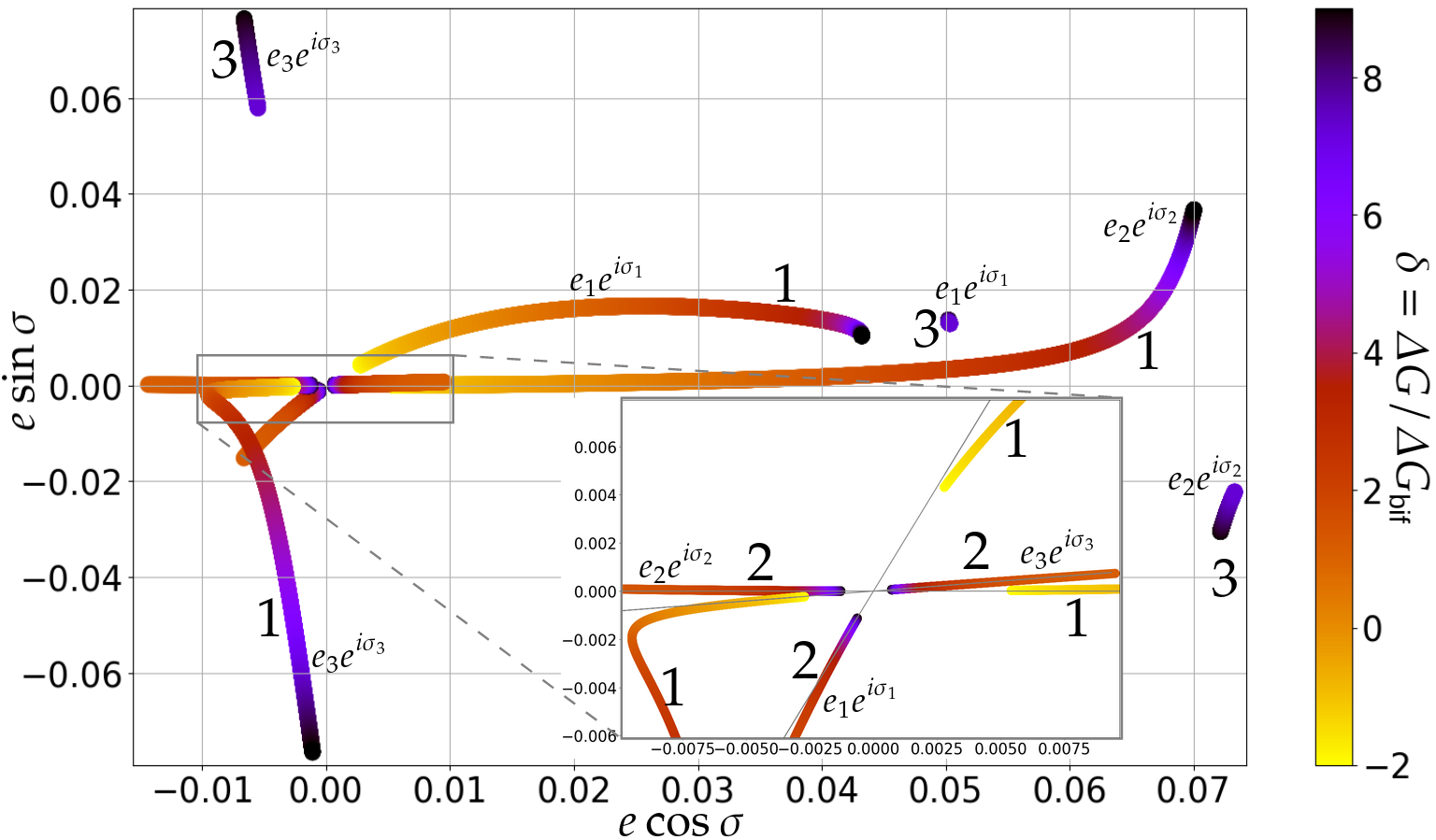}
	\caption{Position of the elliptic branches $1$, $2$ and $3$ (see Table \ref{tableau_de_pf}) of equilibria of the Hamiltonian (\ref{total_hamiltonian}) in the resonance $1:1:2$, for $-2\leq\delta\leq9$. For each branch, three curves appear, corresponding to $e_je^{i\sigma_j}$ for $j = 1,2,3$. The planetary masses are the same as in Table \ref{tableau_de_pf}. A zoom is made close to the origin. In this area, the analytical position of these equilibria, given by Eqs. (\ref{pf_O1_angle}) and (\ref{poly_deg_3}), is plotted by a thin grey line. The agreement is good at small values of the eccentricity but quickly worsens farther from the origin. In particular, the thin grey lines are straight, since $\sigma_j$ does not depend on $\delta$ in Eq. (\ref{pf_O1_angle}). Branch $1$ exists for all values of $\delta$ and has all colours from yellow to dark purple, while branch $3$ only exists at $\delta>5.997$ and thus only has purple.}\label{fig_O1_O2}\end{figure*}
Limiting the work at first order in eccentricity and forcing a decoupling between $\left(L,\xi \right)$ and $\left(D_j,\sigma_j\right)$ gives analytical expressions of the linearised system\footnote{At least of the equilibria and the eigenvalues, we did not obtain the eigendirections.}, but at the cost of strong approximations. We develop here a Newton-Raphson based algorithm to numerically find the equilibria of the model (\ref{total_hamiltonian}) without these approximations.

The vector field $F_0$, derived from the Hamiltonian (\ref{total_hamiltonian}), depends on the choice of the parameter $\delta$ (through $\Delta G$) and once a fixed point is found for a particular value of $\delta$, we repeat the Newton-Raphson algorithm for slowly varying values of $\delta$ in order to travel along the whole branch. We look for equilibria exploring the parallelepiped in the phase space defined by $\left|u_j\right|<0.08$ and $\left| v_j\right|<0.08$ $\left(e_j^2=C_j(u_j^2+v_j^2)\right)$. We choose $L_0=L^{\star}=\Lambda_1^{\star}/\Gamma^{\star}$ as initial condition of the Newton-Raphson method for $L$ since all equilibria are expected to be close to this value, and so no discretisation is necessary along this axis. In the same way, we only choose $\xi_0\in\left\lbrace\pm\pi/3,\pi\right\rbrace$.

We display in Table \ref{tableau_de_pf} all the equilibria that we have found for $\delta=7$, their hyperbolic or elliptic nature, and the value of $\delta$ that gives birth to the branch. Due to the difficulty of exploring a thinly discretised grid in 6 dimensions, we may have not found all the possible equilibria. We discretised the axes $u_j$ and $v_j$ with only $8$ points, testing $3\times8^6$ initial conditions, all of which converged towards $15$ equilibria. Since the Hamiltonian (\ref{total_hamiltonian}) is invariant by the transformation $(\xi,\sigma_j) \longmapsto (-\xi, -\sigma_j)$, 
fixed points with values of the angles different from $0$ or $\pi$ have a symmetric, hence the $\pm$ and $\mp$ signs in Table \ref{tableau_de_pf} (the upper sign corresponds to a fixed point and the lower sign to its symmetric). This symmetry corresponds to the invariance of the system by a rotation of angle $\pi$ around an axis normal to the total angular momentum.

As is seen in Table \ref{tableau_de_pf}, only the branches $1$, $2$ and $3$ of fixed points can be elliptic, and thus, we focus only on them in the rest of this work. In Fig. \ref{fig_O1_O2}, we plot these branches for values of $\delta$ ranging from $-2$ to $9$. For branches $1$ and $2$, which are predicted by the first order in eccentricity, we also plot them as given by Eqs. (\ref{pf_O1_angle}) and (\ref{poly_deg_3}), for comparison. This section shows how the analytical model is unable to locate the equilibria of the simplified Hamiltonian (\ref{total_hamiltonian}) for $e_j \gtrsim 0.005$ (see Fig. \ref{fig_O1_O2}) and does not even give its topology for $ e_j \gtrsim 0.05 $ (see branches $3$, $5$ and $7$ in Table \ref{tableau_de_pf}, that do not exist at first order in eccentricity). This discrepancy between first and second order in eccentricity was already mentioned by \cite{BeMiFe2006} in the case of the two-planet $1:2$ mean motion resonance.
\begin{figure*}[h]
	\centering
	\includegraphics[width=\linewidth]{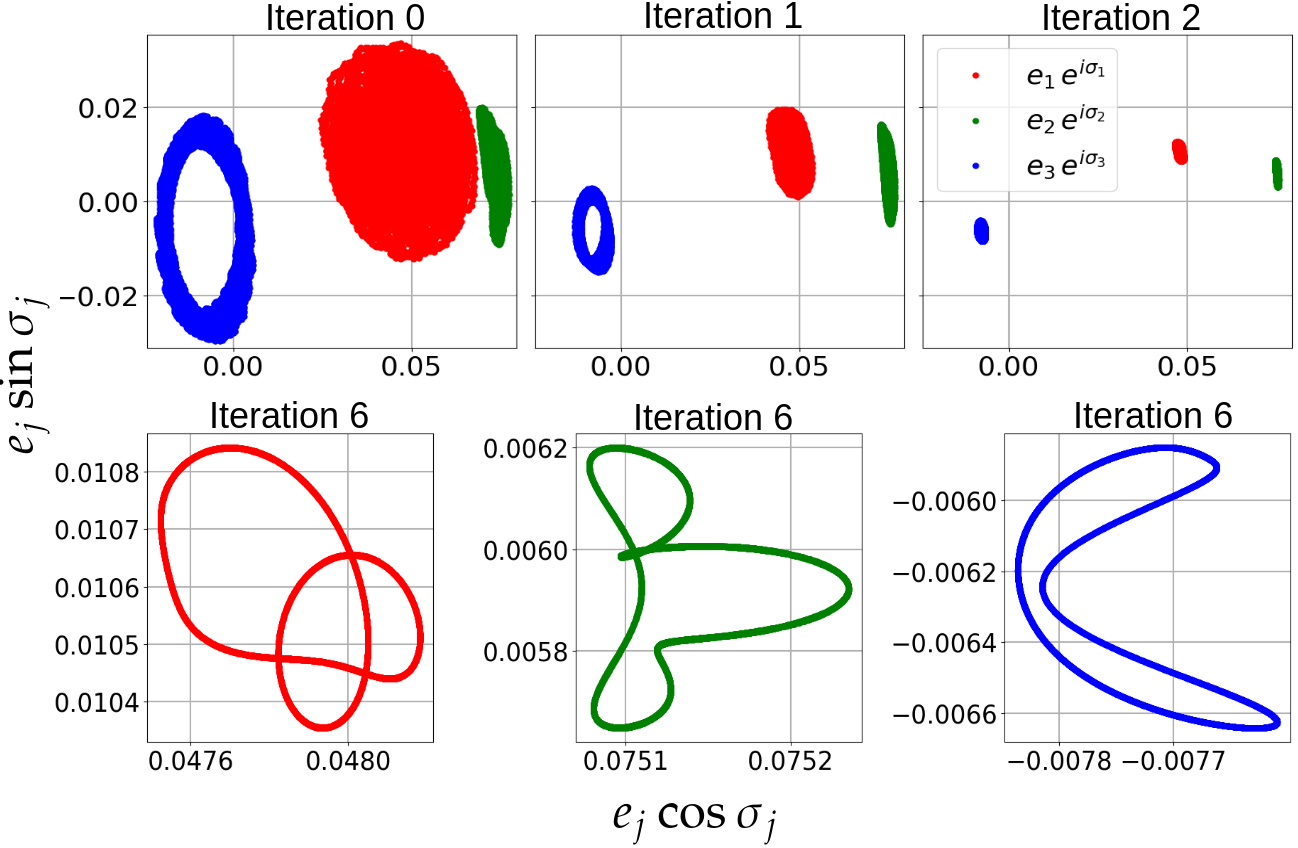}
	\caption{Trajectories of $e_je^{i\sigma_j}$ in the complete Hamiltonian (\ref{complete_hamiltonian}) for the $6$ iterations needed for the algorithm to converge to the \textit{libration centre}. Iteration $0$ is the equilibrium (main branch) of the simplified Hamiltonian  (\ref{total_hamiltonian}) at $\delta=5$. It is rather far from the \textit{libration centre} of the complete Hamiltonian, as is also shown by Fig. \ref{fig_O2_complet}. After $6$ iterations, the algorithm has converged to the \textit{libration centre} and the motion is periodic (hence the closed curves) with frequency $\nu_2$ (see Sect. \ref{pf_vs_centre_libration}). The planetary masses are the same as in Table \ref{tableau_de_pf} and the resonance chain is $1:1:2$.}\label{fig_iter}
\end{figure*}

\subsubsection{Comparison with the complete Hamiltonian}\label{sec_comparison}
In this section, we compare the position of the equilibria of the secular (simplified) Hamiltonian (\ref{total_hamiltonian}) to that of the corresponding periodic orbits of the complete (full) Hamiltonian (\ref{complete_hamiltonian}), that we called \textit{libration centres} in Sect. \ref{pf_vs_centre_libration}. To this aim, we develop an iterative algorithm, similar to what is done by \cite{CoLaCo2010}, to find a \textit{libration centre} of the complete Hamiltonian using an equilibrium of the simplified Hamiltonian as initial condition.
\begin{figure*}[h]
	\centering
	\includegraphics[width=1\linewidth]{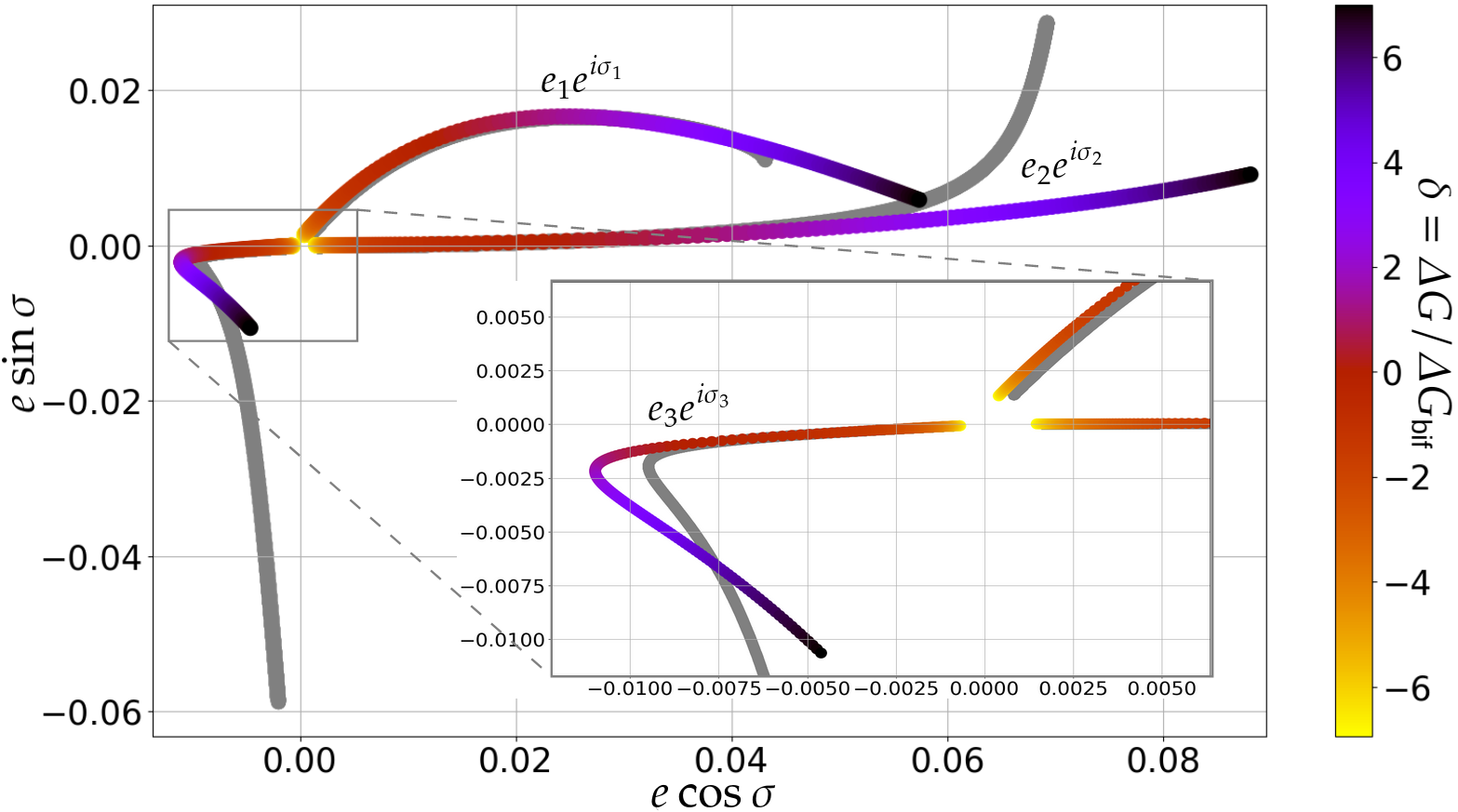}
	\caption{Position of branch $1$ of elliptic \textit{libration centres} of the complete Hamiltonian (\ref{complete_hamiltonian}) in the resonance $1:1:2$ for $-7\leq\delta\leq7$. The planetary masses are the same as in Table \ref{tableau_de_pf}. As comparison, branch $1$ of equilibria of the simplified Hamiltonian (\ref{total_hamiltonian}) is plotted in grey for the same range in $\delta$. For values small enough of $\delta$, the eccentricity is not too high and the agreement is good. The simplified and the complete Hamiltonian diverge when $\delta\rightarrow+\infty$.}\label{fig_O2_complet}
\end{figure*}

We assume that, close enough to a \textit{libration centre} of the complete Hamiltonian (\ref{complete_hamiltonian}), the trajectories are quasiperiodic, and we write, for any complex quantity $z$ depending on these trajectories,
\begin{equation}\label{fourier_serie}
z\left(t\right)=\sum_{\vect{k}\in\mathbb{Z}^6}z_{\vect{k}}e^{i\vect{k}\cdot\vect{\omega}t},
\end{equation}
where the coordinates of $\vect{\omega}=\,^t\left(\nu,\nu_2,\nu_3,g_1,g_2,g_3\right)$ are
the fundamental frequencies of the complete Hamiltonian (\ref{complete_hamiltonian}). The frequencies $\nu$, $\nu_2$ and $\nu_3$ have approximate values given by Eqs. (\ref{nu}), (\ref{nu_23}) and (\ref{nu3}), respectively.

As explained in Sect. \ref{pf_vs_centre_libration}, the \textit{libration centres} correspond to points in the phase space where the motion is periodic\footnote{Quasiperiodic with frequencies $\nu_2$ and $\nu_3$ for a quantity not invariant by rotation around the vertical axis.} with frequency $\nu_2$.
\begin{figure}[h]
	\centering
	\includegraphics[width=1\linewidth]{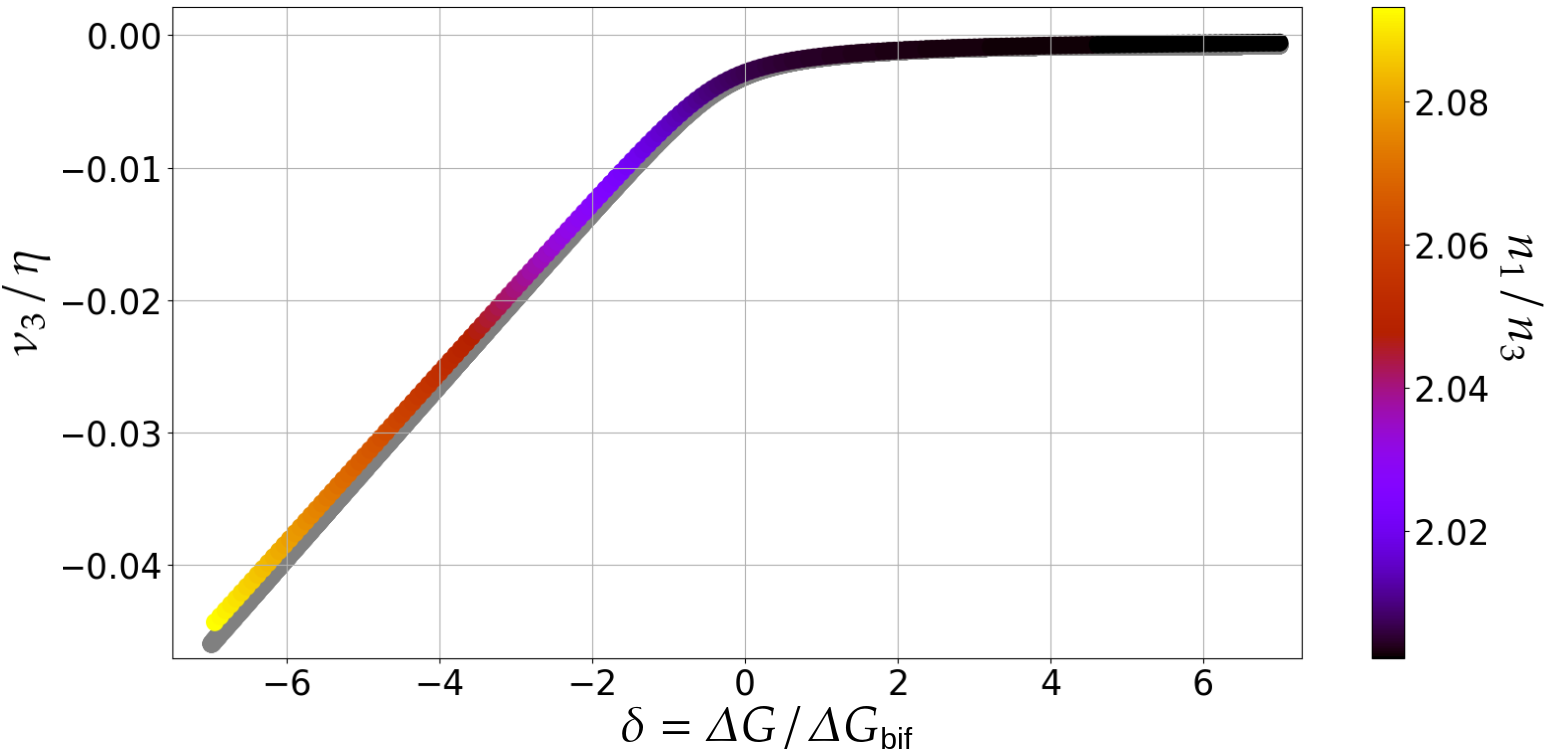}
	\caption{$\nu_3/\eta$ as a function of $\delta$, in the complete Hamiltonian, for the resonance chain $1:1:2$, along the branch $1$ of \textit{libration centres} of Fig. \ref{fig_O2_complet}. The analytical expression (\ref{nu3}) is plotted in grey for comparison. The colour gives the value of $n_1/n_3$. It shows that for high values of $\delta$, this ratio tends towards $\left(p+1\right)/p$ (which is equal to 2 in this case). For small values of $\delta$ though, this ratio diverges from its nominal value and \textit{libration centres} at small values of $\delta$ are far from the Keplerian resonance. For branch $1$, we now refer to very negative values of $\delta$ as far from the resonance and to very positive values of $\delta$ as deep in the resonance. As expected from Eq. (\ref{nu3}), $\nu_3$ is proportionnal to $\delta$ far from the resonance where the eccentricities are small.}\label{fig_nu3_sur_eta}
\end{figure}
The description of the algorithm is as follows:
\begin{itemize}
	\item For a choice of the parameter $\delta$, find the position of an elliptic equilibrium of the simplified Hamiltonian (\ref{total_hamiltonian}) with a Newton-Raphson method. Use it as initial condition to integrate numerically the trajectories of the complete Hamiltonian (\ref{complete_hamiltonian}).
	\item For a complex quantity $z$ depending on the trajectories of (\ref{complete_hamiltonian}), obtain the decomposition (\ref{fourier_serie}) using a frequency analysis method \citep[e.g.][]{Laskar1993}.
	\item Identify terms depending on frequencies other than $\nu_2$ and set to $0$ the corresponding coefficient $z_{\vect{k}}$.
	\item Proceed similarly for different quantities $z$ and evaluate them at time $t=0$ in order to obtain a new initial condition. Restart from the first step using the new initial condition instead of the equilibrium of (\ref{total_hamiltonian}).
	
\end{itemize}
The process is iterated until a convergence occurs. In Fig. \ref{fig_iter}, we display the trajectories of the quantities $e_je^{i\sigma_j}$ in the plane $\left(e_j\cos\sigma_j,\,e_j\sin\sigma_j\right)$, as the algorithm iterates. Isolating terms featuring frequencies other than $\nu_2$ is difficult, if not impossible. Indeed, the frequency analysis gives the scalars $\vect{k}\cdot\vect{\omega}$ but not the vectors $\vect{k}$, which cannot be deduced since the vector $\vect{\omega}$ is unknown for the complete Hamiltonian. We can get around this difficulty by noticing that $\nu_2$ is much larger than the other frequencies, hence it is easy to isolate terms that depend on $\nu_2$ from those which do not. The implementation is thus simplified by setting to $0$ the coefficients $z_{\vect{k}}$ of the terms that do not depend on $\nu_2$.

This algorithm is only able to find \textit{libration centres} associated with elliptic fixed points. We use it to find the branches of \textit{libration centres} associated with branches $1$, $2$ and $3$ of Table \ref{tableau_de_pf}. We confirm the existence of a small hyperbolic zone for branch $2$ in the complete Hamiltonian when the algorithm stops converging as the branch is traveled (by slowly incrementing the value of $\delta$). We plot in Fig. \ref{fig_O2_complet} the main branch (branch $1$) of \textit{libration centres} of the complete Hamiltonian (\ref{complete_hamiltonian}) and we compare it with the main branch of equilibria of the simplified Hamiltonian (\ref{total_hamiltonian}).
In Fig. \ref{fig_nu3_sur_eta}, we plot the precession frequency of the pericentres, $\nu_3$, for the main branch of \textit{libration centres}, and we compare it with the analytical expression (\ref{nu3}). The value of $\nu_3$ in the complete Hamiltonian is obtained from the frequency analysis of $e^{i\xi_3}$, once the \textit{libration centre} is known.

\subsection{Stability map of the $p\!:\!p\!:\!p+1$ resonance chain}\label{sec_map}

\begin{figure*}[h]
	\centering
	\includegraphics[width=1\linewidth]{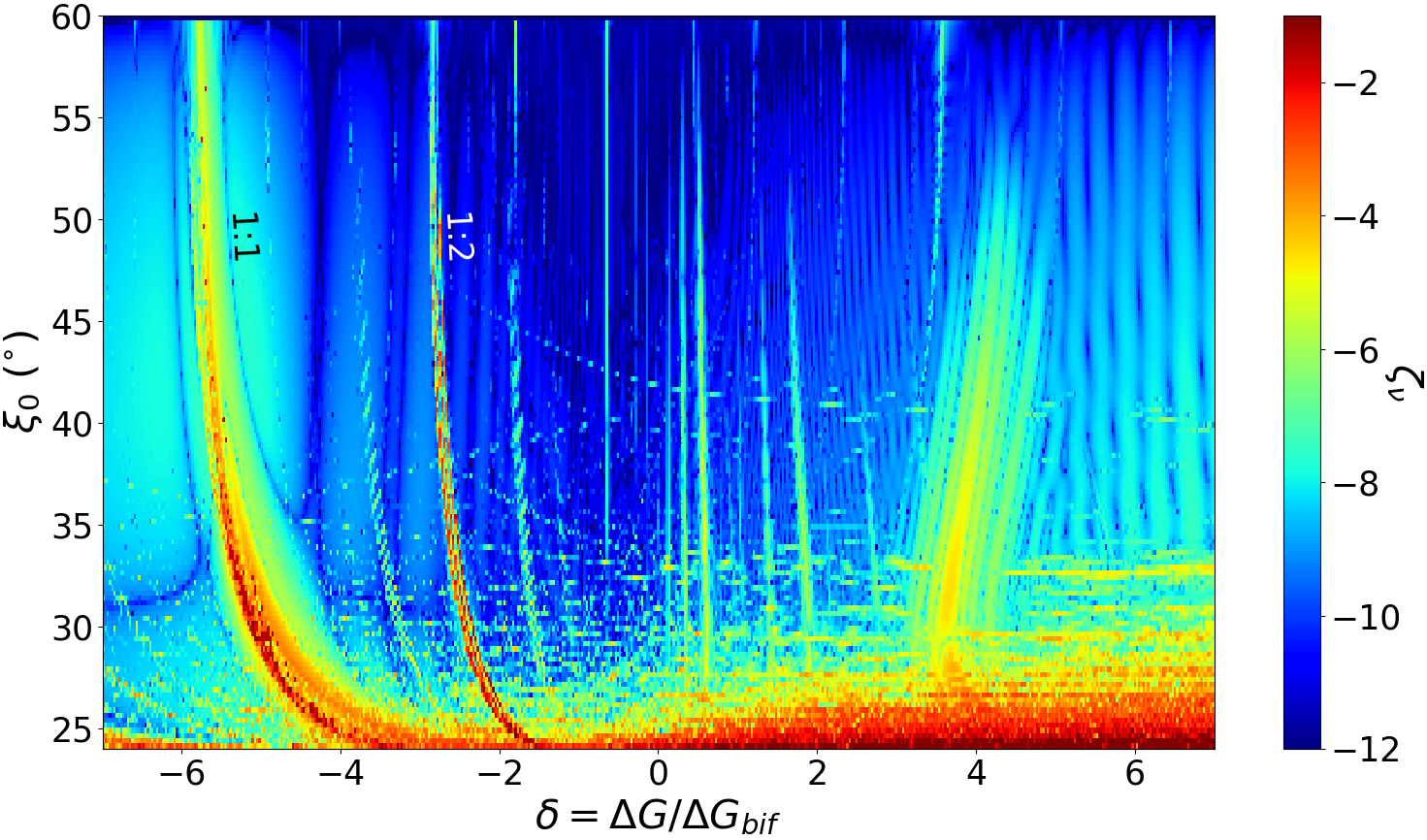}
	\caption{Diffusion index $\zeta_\nu$ as a function of $\delta$ and $\xi_0$. The planetary masses are as in Table \ref{tableau_de_pf} and the chain is $1:1:2$. The top of the figure, at $\xi_0\approx60^{\circ}$, is the main branch. Blue-to-green regions are almost quasi-periodic (stable) while red regions are chaotic (unstable). The secular resonances between the libration frequency $\nu$ and the precession frequency of the pericentres $\nu_3$, predicted by the analytical results (see Fig. \ref{fig_e1_O1+reso_nu_nu3}), are visible, especially the resonance $1:1$, which can lead to chaotic orbits for high enough values of the libration amplitude. Horseshoe-shaped orbits, at the bottom, are mostly chaotic, which is not surprising since $m_1+m_2=2\times10^{-4}$ is close to the limit $3\times10^{-4}$ of their existence \citep{LeleuRobutelCorreia2015}. The main branch around the $1:1$ resonance between $\nu$ and $\nu_3$ is tidally attractive (see Fig. \ref{fig_Re}). Systems undergoing tides entering in this zone can either converge towards the top of the figure or on the contrary become completely chaotic.}\label{fig_map}
\end{figure*}
Before taking into account tidal dissipation in the model, we study the stability of the point-mass $p:p:p+1$ resonance chain by constructing a dynamical map using the frequency analysis method to determine the chaoticity of a given orbit \citep{Laskar1990}. More precisely, we study the stability of the chain along its main branch (see Fig. \ref{fig_e1_O1+reso_nu_nu3}), between $\delta=7$, deep in the resonance, and $\delta=-7$, outside the resonance. 

For each value of $\delta$, we compute the position of the exact \textit{libration centre} by means of the algorithm described in Sect. \ref{sec_comparison}, and we choose an initial value for the angle $\xi$ between its equilibrium value (near $60^{\circ}$) and the value at the boundary between tadpole and horseshoe-shaped orbits \citep[close to $24^{\circ}$, see][]{RobutelPousse2013}. For values of $\xi_0$ close to $60^{\circ}$, the considered orbit is close to the main branch and it moves away for decreasing values of $\xi_0$. For all other variables, we choose as initial condition the value at the \textit{libration centre}. Every trajectory, that is, every choice of $\delta$ and $\xi_0$, is integrated over $80\,000$ periods of the co-orbital planets and for each half of the simulation, the exact value of the libration frequency $\nu$ is extracted from the frequency analysis of $e^{i\xi}$. We obtain two values of $\nu$, namely $\nu^{(1)}$ for the first half of $40\,000$ periods and $\nu^{(2)}$ for the second half. The diffusion index, defined as \citep{RobutelGabern2006}
\begin{equation}
\zeta_\nu=\log_{10}\left|\frac{\nu^{(1)}-\nu^{(2)}}{\nu^{(1)}}\right|,
\end{equation}
measures the degree of quasi-periodicity of the orbit. Orbits with $\zeta_\nu<-6$ are considered close to quasi-periodic (stable) while orbits with $\zeta_\nu>-2$ are very chaotic (unstable). We plot the stability map for the resonance chain $1:1:2$ in Fig. \ref{fig_map}. Secular resonances between $\nu$ and $\nu_3$, already predicted by the analytical results in Fig. \ref{fig_e1_O1+reso_nu_nu3} are visible and induce chaotic motion. Overall, this stability map shows that the chain $p:p:p+1$ is mainly stable.

\section{Tides in the $p:p:p+1$ resonance chain}\label{S3}

In Sect. \ref{S2}, we assumed that the bodies are point mass objects. Here, this approximation is removed and tidal dissipation, due to differential and inelastic deformations of the bodies, is taken into account.

\subsection{Extended Hamiltonian and equations of motion}\label{sec_extended}

Tidal contributions to the orbital evolution of the system follow a very general formulation initiated by \cite{Darwin1880}. Differential interactions between the bodies raise tidal bulges and the subsequent redistribution of mass is responsible for a perturbation in the gravitational potential generated by body $j$ at any point $\vect{r}$ in the space. This perturbation is given by \citep[e.g.][]{Kaula1964}
\begin{equation}
V_{i,j}(\vect{r})=-\kappa_2^{(j)}\frac{\mathcal{G}m_i}{R_j}\left(\frac{R_j}{r}\right)^3\left(\frac{R_j}{r_i^{\bigstar}}\right)^3 P_2\left(\cos S\right),
\end{equation}
where the indice $i$ (resp. $j$) refers to the body responsible for the tidal bulge (resp. where the bulge is raised), $\vect{r}_i$ is the position of body $i$ with respect to the barycenter of body $j$, $R_j$ is the radius of the body $j$, $\kappa_2^{(j)}$ is its second Love number, $P_2$ is the second Legendre polynomial, and $S$ is the angle between $\vect{r}_i^{\bigstar}$ and $\vect{r}$. For a body of mass $m_k$, located at $\vect{r}_k$ and interacting with this bulge, the increment in potential energy is
\begin{equation}\label{total_potential}
U_{i,j,k}\left(\vect{r}_k\right)=m_kV_{i,j}\left(\vect{r}_k\right).
\end{equation}
For $N=4$ tidally interacting bodies, $N\left(N-1\right)^2=36$ such potentials are generated. In the case of planets orbiting a Solar type star though, only tides raised by the star on the planets and felt by the star have to be considered, since they are dominant with respect to any other contribution \citep[see][]{CoRoCo2021}. That is, we only consider the three contributions $U_{i,j,k}$ with $i=k=0$ and $j\in\left\lbrace1,2,3\right\rbrace$.

The dissipation of mechanical energy inside the planets introduces a time delay $\Delta t$ between the tidal stress and the corresponding deformation. As a consequence, the tidal bulge is not aligned with the star and the subsequent torque affects the spins and orbits of the planets. For any quantity $z_j$ related to planet $j$, we note
\begin{equation}\label{star}
z_j^{\bigstar}\left(t\right) = z_j\left(t-\Delta t_j\right). 
\end{equation}
For a frequency of excitation $\stigma$, the quality factor $Q_j\left(\stigma\right)$ \citep{Munk1960}, which measures the amount of energy dissipated in a period $2\pi/\stigma$, is related to the time delay by
\begin{equation}
Q_j^{-1} (\stigma) = \sin ( \stigma \Delta t_j (\stigma) ) \approx \stigma \Delta t_j (\stigma).
\end{equation}
The dependency of $\Delta t_j (\stigma)$ on $\stigma$ is unknown and a simple, commonly used rheology, consists in considering that the time delay is independent on the frequency \citep{Mignard1979}. We adopt this tidal model in this work, reducing the rheology to the constant parameters $\kappa_2^{(j)}$ and $Q_j^{-1}=n_{j,0}\Delta t_j$.

Although tides do not preserve the total energy, the Hamiltonian formalism is extended by considering the starred variables as parameters when deriving the equations of motion. Their contribution to the Hamiltonian reads
\begin{equation}
H_t=\sum_{j\leq 3}\left(U_t^{(j)}+T_j\right),
\end{equation}
where, in the heliocentric reference frame
\begin{equation}
U_t^{(j)}=-\kappa_2^{(j)}\mathcal{G}m_0^2\frac{R_j^5}{r_j^3r_j^{\bigstar3}}P_2\left(\cos S\right),\;\;T_j=\frac{{\Theta'_j}^2}{2\alpha_jm_jR_j^2},
\end{equation}
with
\begin{equation}
S=\lambda_j-\lambda_j^{\bigstar}-\left(\theta_j-\theta_j^{\bigstar}\right),
\end{equation}
and $\theta_j$ is the rotation angle of body $j$, $\omega'_j=d\theta_j/dt$ is its rotation rate, $\Theta'_j=\alpha_jm_jR_j^2\omega'_j$ is the conjugated momentum of $\theta_j$, and $\alpha_j$ is a dimensionless structure constant depending on the state equation of body $j$, such that $\alpha_jm_jR_j^2$ is its principal moment of inertia. The transformations (\ref{change_var_action}) and (\ref{normalization}) are performed on the tidal Hamiltonian, with the normalisations $\Theta_j=\Theta'_j/\Gamma^{\star}$ and $\mathcal{T}_j=T_j/\Gamma^{\star}$. Denoting
\begin{equation}
\begin{split}
&q_j=\kappa_2^{(j)}\qoppa_j^5,\;\;\qoppa_j=\frac{R_j}{a_{j,0}},\;\;\R_j=C_j\Lambda_j=\frac{\tilde{\Lambda}_j}{\Lambda_j^{\star}}\approx 1\\&\text{and}\;\;\Delta\vartheta=\vartheta-\vartheta^{\bigstar},
\end{split}
\end{equation}
where $\vartheta$ stands for any angle, we obtain for the tidal Hamiltonian
\begin{figure*}[h]
	\centering
	\includegraphics[width=1\linewidth]{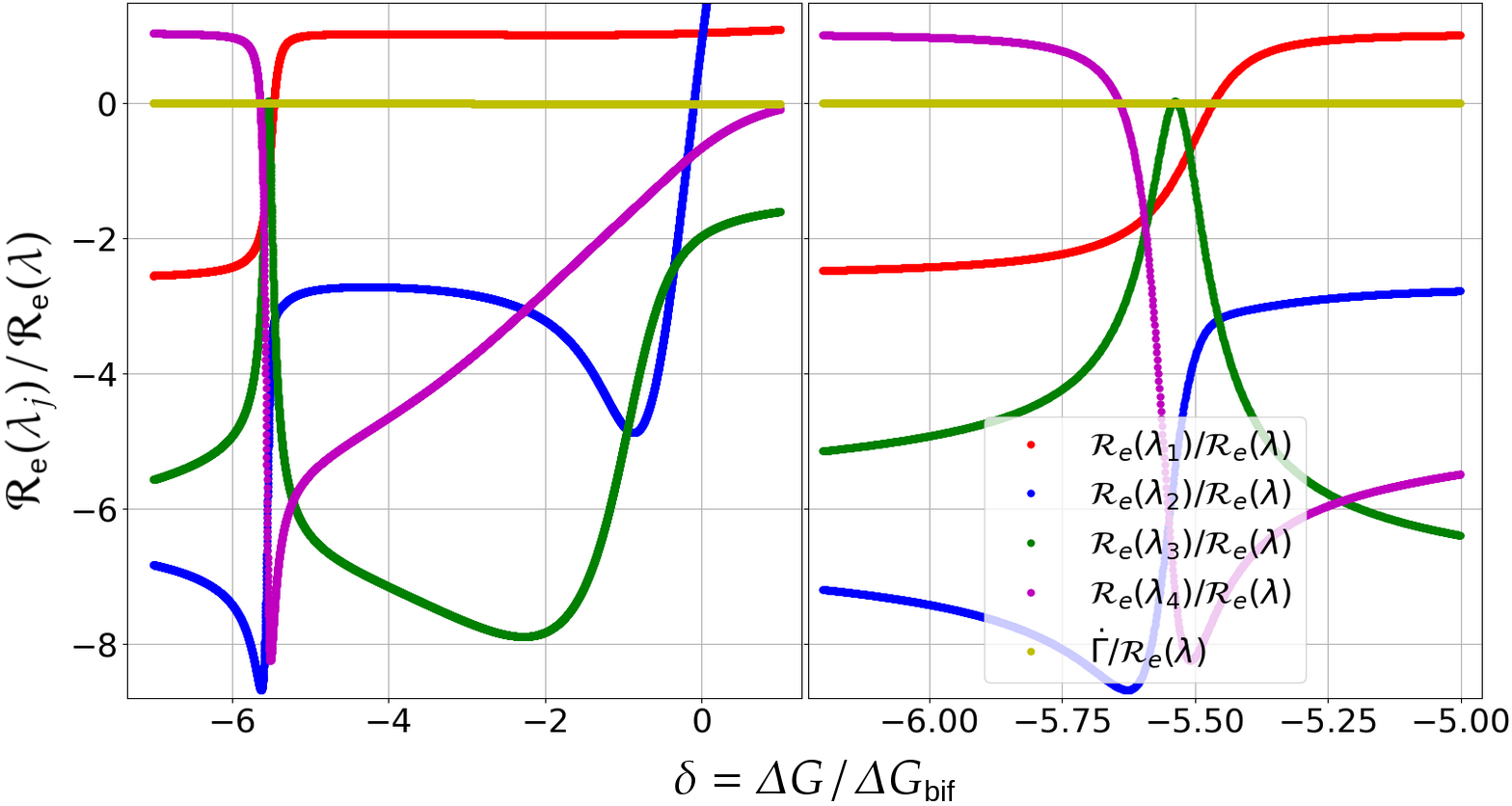}
	\caption{Real parts of the eigenvalues of the linearised system associated to $F$ in the vicinity of its main branch of pseudo equilibria, for $-7\leq\delta\leq1$ (left) and $-6.2\leq\delta\leq-5$ (right). The planetary masses are as in Table \ref{tableau_de_pf} and the resonance chain is $1:1:2$. Only the $4$ eigenvalues associated with the four degrees of freedom of the conservative system are represented. The other eigenvalues (e.g. associated with the rotation rates $\omega_j$) are of no interest. The real parts behave erratically at the $1:1$ resonance between $\nu$ and $\nu_3$ (see Fig. \ref{fig_e1_O1+reso_nu_nu3}), in such a way that all of them are negative for $-5.64\leq\delta\leq-5.47$. We thus expect the main branch to be linearly stable in this region. For $\delta\leq-0.104$, at least $3$ out of $4$ real parts are negative. In the region $\delta\leq1$, we have $|\dot{\Gamma}|/\R_{\text{e}}\left(\lambda\right)\leq0.012$, and the criterion (\ref{criterium}) is very well respected. Similar figures with other planetary masses are available in appendix \ref{append_simu}.}\label{fig_Re}
\end{figure*}
\begin{equation}
\Ham_t=\sum_{j\leq 3}\left(\mathcal{U}_t^{(j)}+\mathcal{T}_j\right),
\end{equation}
with \citep{CoRoCo2021}
\begin{equation}
\begin{split}
&\mathcal{U}_t^{(j)}=-q_j\frac{m_0}{m_j}n_{j,0}C_j^{-1}\R_j^{-6}\R_j^{\bigstar-6}\left(A_t^{(j)}+\Xi_j\right),\\&\mathcal{T}_j=\frac{C_jn_{j,0}\Theta_j^2}{2\alpha_j\qoppa_j^2},\\
&\Xi_j=B_t^{(j)}\left(\R_j^{-1}D_j+\R_j^{\bigstar-1}D_j^{\bigstar}\right)\\&+\left(\R_j\R_j^{\bigstar}\right)^{-1/2}\R_{\text{e}}\left(C_t^{(j)}\sqrt{D_jD_j^{\bigstar}}e^{i\Delta\varpi_j}\right)
\end{split}
\end{equation}
and
\begin{equation}
\begin{split}
&A_t^{(j)}=\frac{1}{4}+\frac{3}{4}\cos 2\left(\Delta\lambda_j-\Delta\theta_j\right),\\&B_t^{(j)}=\frac{3}{4}-\frac{15}{4}\cos 2\left( \Delta\lambda_j-\Delta\theta_j \right),\\
&C_t^{(j)}\!=\!\frac{3}{8}e^{i\left( \Delta\lambda_j-2\Delta\theta_j \right)}\!+\!\frac{9}{4}e^{-i\Delta\lambda_j}\!+\!\frac{147}{8}e^{-i\left( 3\Delta\lambda_j-2\Delta\theta_j\right)}.\!\!\!\!\!
\end{split}
\end{equation}
Note that no expansion at order $0$ in the vicinity of $\Lambda_j^{\star}$ is performed since it loses relevant tidal dynamics \citep[see][]{CoRoCo2021}. We instead keep exact expressions in $\Lambda_j$. The differential system is derived from the tidal Hamiltonian using the Hamilton-Jacobi equations and considering the starred variables as parameters. The starred variables are then expressed by a first order Taylor expansion in Eq. (\ref{star}). We refer the reader to \cite{CoRoCo2021} for more details. The perturbation to the vector field, due to tides and at second order in eccentricity, reads
\begin{equation}\label{vector_field_tides}
\begin{split}
&\dot{D}_j=3n_{j,0}D_j\frac{m_0}{m_j}\R_j^{-13}\frac{q_j}{Q_j}\left( 12\omega_j+57\R_j-76\right),\\
&\dot{L}\!=\!3\eta\frac{m_0}{m_1}\R_1^{-13}\frac{q_1}{Q_1}\!\left\lbrace\omega_1\!\left(\Lambda_1\!+\!27D_1\right)\!+\!\left(3\R_1\!-\!4\right)\!\left(\Lambda_1\!+\!46D_1\right)\right\rbrace\!,\!\!\!\!\!\!\!\!\!\!\!\!\!\!\!\!\!\!\!\!\!\!\!\!\!\!\!\!\!\!\\
&\dot{\sigma}_j=-\frac{15}{2}n_{j,0}\frac{m_0}{m_j}q_j\R_j^{-13},\\
&\dot{\xi}=\frac{3}{2}\eta\frac{m_0}{m_1}q_1\R_1^{-13}\frac{4\Lambda_1\!+\!65D_1}{\Lambda_1}\!-\!\frac{3}{2}\eta\frac{m_0}{m_2}q_2\R_2^{-13}\frac{4\Lambda_2\!+\!65D_2}{\Lambda_2},\!\!\!\!\!\!\!\!\!\!\!\!\!\!\!\!\!\!\!\!\!\!\!\!\!\\
&\dot{\Gamma}=\sum_{j\leq 3}3\frac{n_{j,0}}{n_{3,0}}n_{j,0}\frac{m_0}{m_j}\R_j^{-13}\frac{q_j}{Q_j}\\
&\place\;\;\;\;\;\;\left\lbrace\omega_j\left(\Lambda_j+27D_j\right)+\left(3\R_j-4\right)\left(\Lambda_j+46D_j\right)\right\rbrace,\\
&\dot{g}=\sum_{j\leq 3}3n_{j,0}\frac{m_0}{m_j}\R_j^{-13}\frac{q_j}{Q_j}\\
&\place\;\;\;\;\;\;\left\lbrace\omega_j\left(\Lambda_j+15D_j\right)+\left(3\R_j-4\right)\left(\Lambda_j+27D_j\right)\right\rbrace,\\
&\dot{\omega}_j=-3n_{j,0}C_j\alpha_j^{-1}\qoppa_j^{-2}\frac{m_0}{m_j}\R_j^{-13}\frac{q_j}{Q_j}\\
&\place\;\;\;\;\;\;\left\lbrace\omega_j\left(\Lambda_j+15D_j\right)+\left(3\R_j-4\right)\left(\Lambda_j+27D_j\right)\right\rbrace,
\end{split}
\end{equation}
where we posed $\omega_j=\omega'_j/n_{j,0}$. Since the differential system does not depend on $\xi_2$ and $\xi_3$ and as their dynamics is of no interest to us, the lines $\dot{\xi}_2$ and $\dot{\xi}_3$ are absent from the differential system (\ref{vector_field_tides}).

\subsection{Pseudo-fixed points and linearisation in their vicinity}\label{sec_pseudo_fixed}

The total differential system that we consider for our model is the one derived from the Hamiltonian (\ref{total_hamiltonian}), noted $F_0$, to which we now add the tidal perturbations (\ref{vector_field_tides}). We note it $F:\mathbb{R}^{13}\mapsto\mathbb{R}^{13}$. We want here to find the equilibria of $F$ and to study the linearised dynamics in their vicinity. However, although $F_0$ has equilibria (see Table \ref{tableau_de_pf}), $F$ has none. Indeed, the five lines of (\ref{vector_field_tides}) corresponding to $\dot{\Gamma}$, $\dot{g}$, $\dot{\omega}_1$, $\dot{\omega}_2$ and $\dot{\omega}_3$ cannot all vanish if $D_j\neq0$. Since $F_0$ has no equilibria at $D_j=0$ and does not contribute to these five lines, we conclude that $F$ has no equilibria. If the planets are all synchronised, that is, if the $\dot{\omega}_j$ all vanish\footnote{It implies that $\dot{g}$ also vanishes, since $\dot{g}+\sum\alpha_j\qoppa_j^2C_j^{-1}\dot{\omega}_j=0$ by conservation of the total angular momentum.}, then $\dot{\Gamma}<0$ and
\begin{equation}\label{gamma_drift}
\dot{\Gamma}\approx-\sum_{j\leq 3}21\frac{n_{j,0}}{n_{3,0}}n_{j,0}\frac{m_0}{m_j}\frac{q_j}{Q_j}D_j\propto-\sum_{j\leq 3}\frac{m_0}{m_j}\frac{q_j}{Q_j}e_j^2.
\end{equation}
We call pseudo-equilibrium of $F$, or pseudo-fixed point of $F$, a point $X\in\mathbb{R}^{13}$ such that $F\left(X\right)=\,^t\left(0,0,0,0,0,0,0,0,\dot{\Gamma}\left(X\right),0,0,0,0\right)$. Even though it does not have equilibria, $F$ has pseudo-equilibria and we find them using an extension of the Newton-Raphson based algorithm that we developed in Sect. \ref{sec_topology}.

Eq. (\ref{gamma_drift}) shows that on a branch of pseudo equilibria of $F$, the parameter $\Gamma$ (and thus the parameter $\delta$, see Eq. (\ref{Deltag_def})) drifts at a speed proportionnal to the square of the eccentricities. This means that, with tides, the main branch is traveled from right to left on Fig. \ref{fig_e1_O1+reso_nu_nu3} \citep{DeLaCo2014}, much quicker when $\delta>0$ than when $\delta<0$ (due to high eccentricities for positive $\delta$). As the branch is traveled, whether or not the system stays close to it or moves away depends on the linear stability of the differential system $F$ in the vicinity of the branch. That is, it depends on the real parts of the eigenvalues of the linear system associated to $F$. Since $\Gamma$ is not constant at the pseudo-fixed points but drifts at a speed given by Eq. (\ref{gamma_drift}), computing the eigenvalues of the linearised system makes sense only if $\Gamma$ drifts slowly enough, that is, only if 
\begin{equation}\label{criterium}
\big|\dot{\Gamma}\big|\ll\max_{k\leq 13}\left|\R_{\text{e}}\left(\lambda_k\right)\right|,
\end{equation}
where the $\lambda_k$ are the eigenvalues of the linearised system. Indeed, $|\dot{\Gamma}|^{-1}$ is the timescale of evolution of $\Gamma$ while $\left(\max_{k\leq13}\left|\R_{\text{e}}\left(\lambda_k\right)\right|\right)^{-1}$ is the timescale of tidal evolution. When the criterion (\ref{criterium}) is fulfilled, $\Gamma$ can be considered constant on the timescale of tidal evolution, and the real parts of the linearised system have physical meaning.

Branch $3$ always has high eccentricities (see Fig. \ref{fig_O1_O2}) and exists only for $\delta>5.997$. The drift in $\delta$ towards negative values is quick at high eccentricity (see Eq. (\ref{gamma_drift})), and so, branch $3$ is tidally very unstable and uninteresting to us. Branch $2$ has small values of the eccentricities at large $\delta$ but the existence of an hyperbolic zone at $5.55\leq\delta\leq5.80$ (see Table \ref{tableau_de_pf}) makes it uninteresting too, since the drift ensures that this zone is reached. Hence, we limit the study of tidal dissipation to the main branch (branch $1$).

\begin{figure}[h]
	\centering
	\includegraphics[width=1\linewidth]{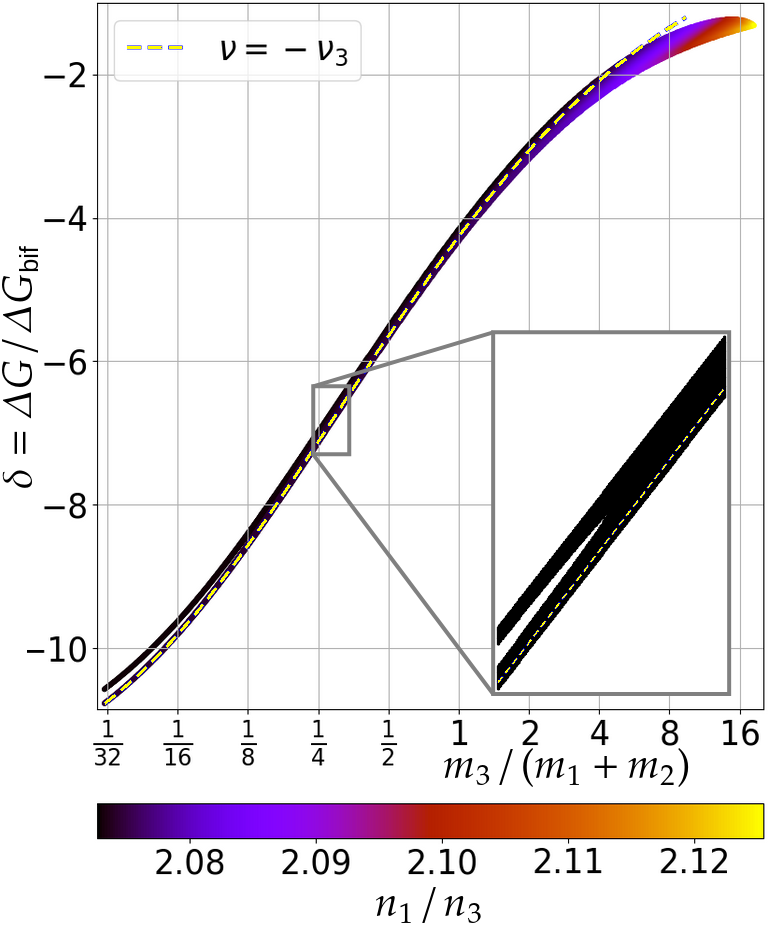}
	\caption{Position of the \textit{linearly stable region} for the resonance chain $1:1:2$ in the plane $\left(m_3,\delta\right)$. For every point in this plane, we computed the eigenvalues of the linearised system associated to the vector field $F$ and plotted the point only if all the real parts are negative. The co-orbital masses are $m_1=m_2=10^{-4}$ and the tidal parameters are those of the system $0$ in Table \ref{simu_param}. The position of the \textit{linearly stable region} weakly depends on $m_1/m_2$ and on the tidal parameters. The colour gives $n_1/n_3$ and shows that the chain stabilises the dynamics far from the Keplerian resonance (for which $n_1/n_3=2$). The dashed yellow line plots the secular $1:1$ resonance between $\nu$ and $\nu_3$, computed with Eqs. (\ref{nu}) and (\ref{nu3}), respectively. For $m_3>18\left(m_1+m_2\right)$, the \textit{linearly stable region} disappears, while for $m_3<0.29\left(m_1+m_2\right)$, two distinct \textit{linearly stable regions} exist, whose widths tend to $0$ with $m_3$. We discuss the impact of $m_3$ on the dynamics in appendix \ref{append_simu}.}\label{fig_lsr}
\end{figure}

In Fig.~\ref{fig_Re}, we plot the real parts of the eigenvalues of the linearised system associated to $F$, along its main branch of pseudo equilibria, which is a little perturbation of the main branch of equilibria of (\ref{total_hamiltonian}). To guarantee that the condition (\ref{criterium}) is well respected, we limit ourselves to $\delta<1$. This is not really a restriction, since tides ensure that this region is
quickly reached. We also plot in the same figure the value of $\dot{\Gamma}$ for comparison. Only the eigenvalues which are the perturbations of (\ref{chi}) and (\ref{nu}) are plotted. In the absence of a third planet, \cite{CoRoCo2021} have shown that the eigenvalue responsible for the exponential increase of the libration amplitude of $\xi$, and thus for the destruction of the co-orbital motion, has a real part
\begin{equation}\label{Re_lambda}
\R_{\text{e}}\left(\lambda\right)=\frac{9}{2}\eta\frac{m_0}{m_1+m_2}\left(\frac{m_1}{m_2}\frac{q_2}{Q_2}+\frac{m_2}{m_1}\frac{q_1}{Q_1}\right),
\end{equation}
and we normalise by this quantity in Fig. \ref{fig_Re} and appendix \ref{append_simu}. The region $-5.64\leq\delta\leq-5.47$, at the $1:1$ secular resonance between the libration frequency of the co-orbitals, $\nu$, and the frequency of all the pericentres at the pseudo equilibria, $\nu_3$, is such that all the eigenvalues of the system have negative real parts, and we expect this region to be linearly stable\footnote{Even though it is slightly chaotic along the main branch for the conservative system, see the map of Fig. \ref{fig_map}.}. We show in Sect. \ref{S4} that this is indeed the case. The linear
stability is only temporary though, since the drift in $\delta$ ensures that this region is eventually left. We call this region \textit{linearly stable region} in the remaining of this work. The range in $\delta$ corresponding to the \textit{linearly stable region} strongly depends on $m_3/\left(m_1+m_2\right)$. In Fig. \ref{fig_lsr}, we display its position in the plane $\left(m_3,\delta\right)$.

\section{Numerical simulations and discussions}\label{S4}

In this section, we investigate the ability of our model to predict the behaviour of a system in the $p:p:p+1$ resonance under tidal dissipation, while we also check the results drawn in Sect. \ref{sec_pseudo_fixed} on the linearised dynamics in the vicinity of the main branch.
\subsection{Procedure}
We numerically integrate two different sets of equations. The first set, our model, is the differential system $F$, that is, the vector field derived from (\ref{total_hamiltonian}) to which we add the tidal perturbations (\ref{vector_field_tides}). The second set is a direct $N$-body simulation of the complete system, with the constant-$\Delta t$ model, given by the set of Eqs. (\ref{nbody_direct}).

When the third planet is absent, the relevant parameters to consider to predict the destruction time of a system of two co-orbital planets are \citep{CoRoCo2021}
\begin{equation}\label{param_tides_def}
\Omega=\frac{q_1}{Q_1}+\frac{q_2}{Q_2},\; x=\frac{m_1}{m_2},\; y=\frac{q_2Q_1}{q_1Q_2},\;\varepsilon=\frac{m_1+m_2}{m_0},
\end{equation}
namely, the total dissipation rate, the mass ratio, the dissipation rate ratio, and the total co-orbital mass ratio. Since we are interested in comparing the lifetime of the co-orbitals when they are inside the resonance chain $1:1:2$ with their lifetime when they are alone, we make use of these parameters. Due to the complexity of the dynamics of the resonance chain $p:p:p+1$, we do not have analytical
\begin{figure*}[h]
	\centering
	\includegraphics[width=1\linewidth]{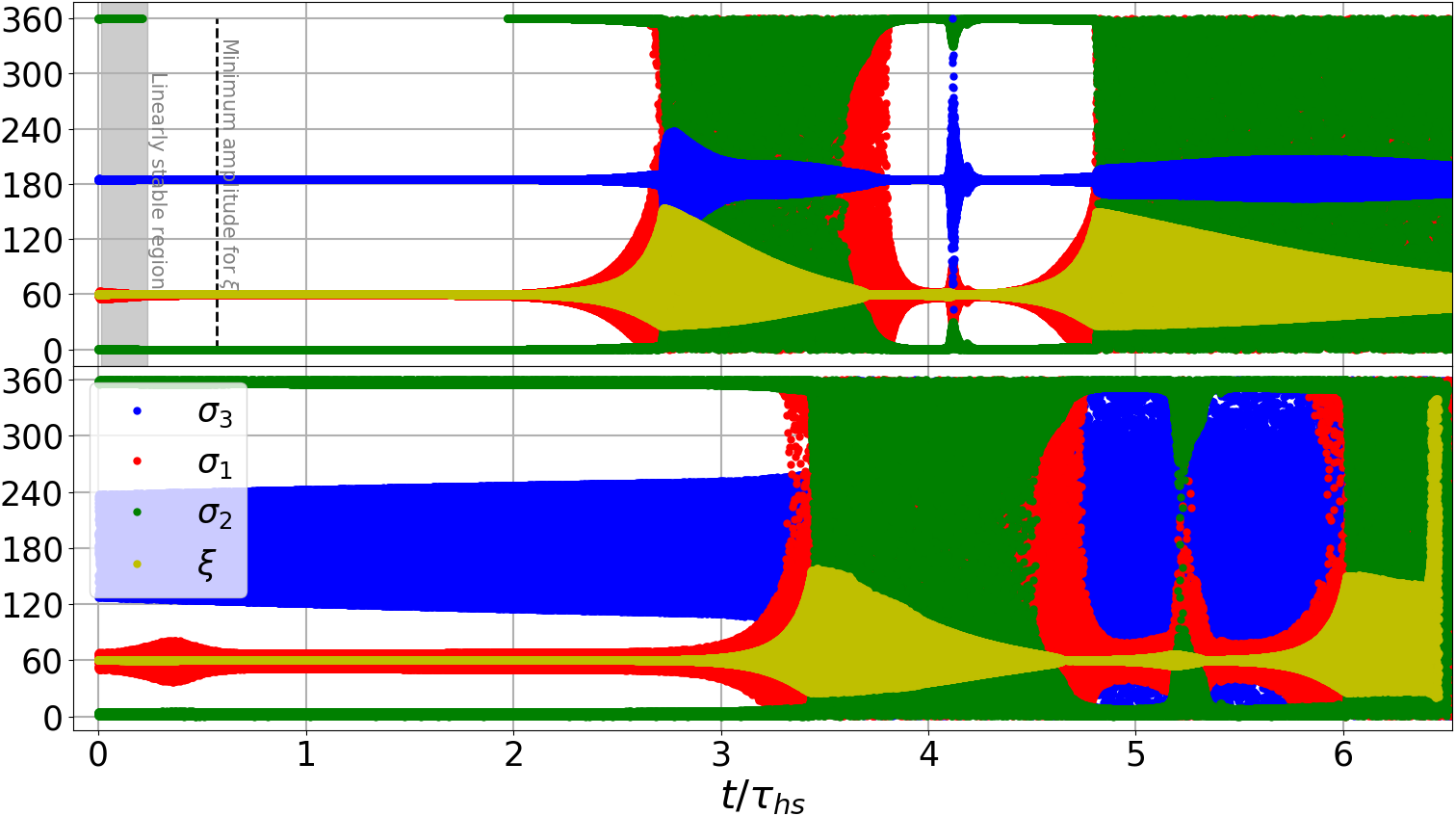}
	\caption{Evolution of $\xi$ and the $\sigma_j$ (in degrees) as a function of time for system $1$ (see Table \ref{simu_param}) as integrated by the simplified model $F$ (Eqs.\,(\ref{total_hamiltonian}) and (\ref{vector_field_tides})) (top) and the direct $N$-body simulation (\ref{nbody_direct}) (bottom). In the direct simulation, horseshoe-shaped orbits are reached after $6.42\,\tau_{\text{hs}}$, and the co-orbital motion is destroyed shortly after that, as expected from the stability map (Fig.~\ref{fig_map}). The model reaches the horseshoe-shaped orbits at $t=13.1\,\tau_{\text{hs}}$. The negative real parts of all the eigenvalues in the grey-shaded area allow the libration amplitude of $\xi$ to reach values as small as $1.8$ arc seconds at $t=0.57\,\tau_{\text{hs}}$. This minimum happens after the \textit{linearly stable region} is left, since the proper mode associated to the newly positive real part (the purple one in Fig. \ref{fig_Re}) had been completely damped by the \textit{linearly stable region} and some time is needed to pump it noticeably. For this choice of masses and $\delta$, the presence of the chain increases the lifetime of the co-orbitals by a factor $6.42$. The thickness of the lines on the bottom plot (see e.g. $\sigma_3$) is due to the short-period oscillations that were averaged out in the model.}\label{fig_angle_1}
\end{figure*}
expressions depending on the parameters of the lifetime of the system, and trying to draw a complete picture would require a very large number of simulations. We are rather interested in performing a small number of simulations with parameters that we judge interesting. Thus, we only show the evolution of two systems for the chain $1:1:2$, whose parameters are given in Table \ref{simu_param}.
\begin{table}[b]
	\begin{center}
		\begin{tabular}{rrrrrr}
			\hline
			\#&$\Omega$&$x$&$y$&$\varepsilon$&$\delta$\\
			\hline
			$0$&$4\times10^{-12}$&$1$&$1$&$2\times10^{-4}$&$-5.51$\\
			$1$&$4\times10^{-12}$&$1/10$&$100$&$2\times10^{-4}$&$-5.46$\\
			\hline
		\end{tabular}
		\caption{Parameters of the two numerical simulations.}\label{simu_param}
	\end{center}
	{\small In this table, the chosen value of $\delta$ is that of the maximum of the region where all the real parts are negative, ensuring that the system crosses all the \textit{linearly stable region}. The rotations of the planets are initially synchronised and both sets of parameters verify $m_3/m_0=10^{-4}$, $\kappa_{2}^{(3)}=0$ and $\bar{a}=0.02$ AU. System $1$ is the system that was used for all the figures in Sects. \ref{S2} and \ref{S3}.}
\end{table}
Nevertheless, we performed additional simulations with different choices for the planetary masses and the initial $\delta$. The most interesting ones are presented in appendix \ref{append_simu}, where we thoroughly discuss the influence of a larger or smaller value for $m_3$.

For the systems listed in Table~\ref{simu_param}, the chosen value of $\delta$ is such that the beginning of the simulation is at the rightmost point of the \textit{linearly stable region} (these regions are $-5.67\leq\delta\leq-5.52$ for system $0$ and $-5.64\leq\delta\leq-5.47$ for system $1$, see Fig. \ref{fig_Re}). These systems are thus expected to be initially very stable, until they leave this region (due to the drift in $\delta$, see Sect. \ref{sec_pseudo_fixed}). 

To integrate both systems with the simplified model $F$ (Eqs.\,(\ref{total_hamiltonian}) and (\ref{vector_field_tides})), we find, for the given value of $\delta$ and the planetary masses, the position of the fixed point of the Hamiltonian (\ref{total_hamiltonian}), and use it as initial condition for the integration, with a shift $\Delta \xi=0.1^{\circ}$ in $\xi$, in order to not start exactly at the fixed point. The pseudo fixed point of the model with tides is very close to the fixed point of (\ref{total_hamiltonian}), and we ignore the difference. To integrate the system with the direct $N$-body set of Eqs. (\ref{nbody_direct}), we find, for the given value of $\delta$ and the planetary masses, the position of the \textit{libration centre} with the algorithm described in Sect. \ref{sec_comparison}, and use it as initial condition for the integration, again with the shift $\Delta \xi=0.1^{\circ}$.
\begin{figure*}[h]
	\centering
	\includegraphics[width=1\linewidth]{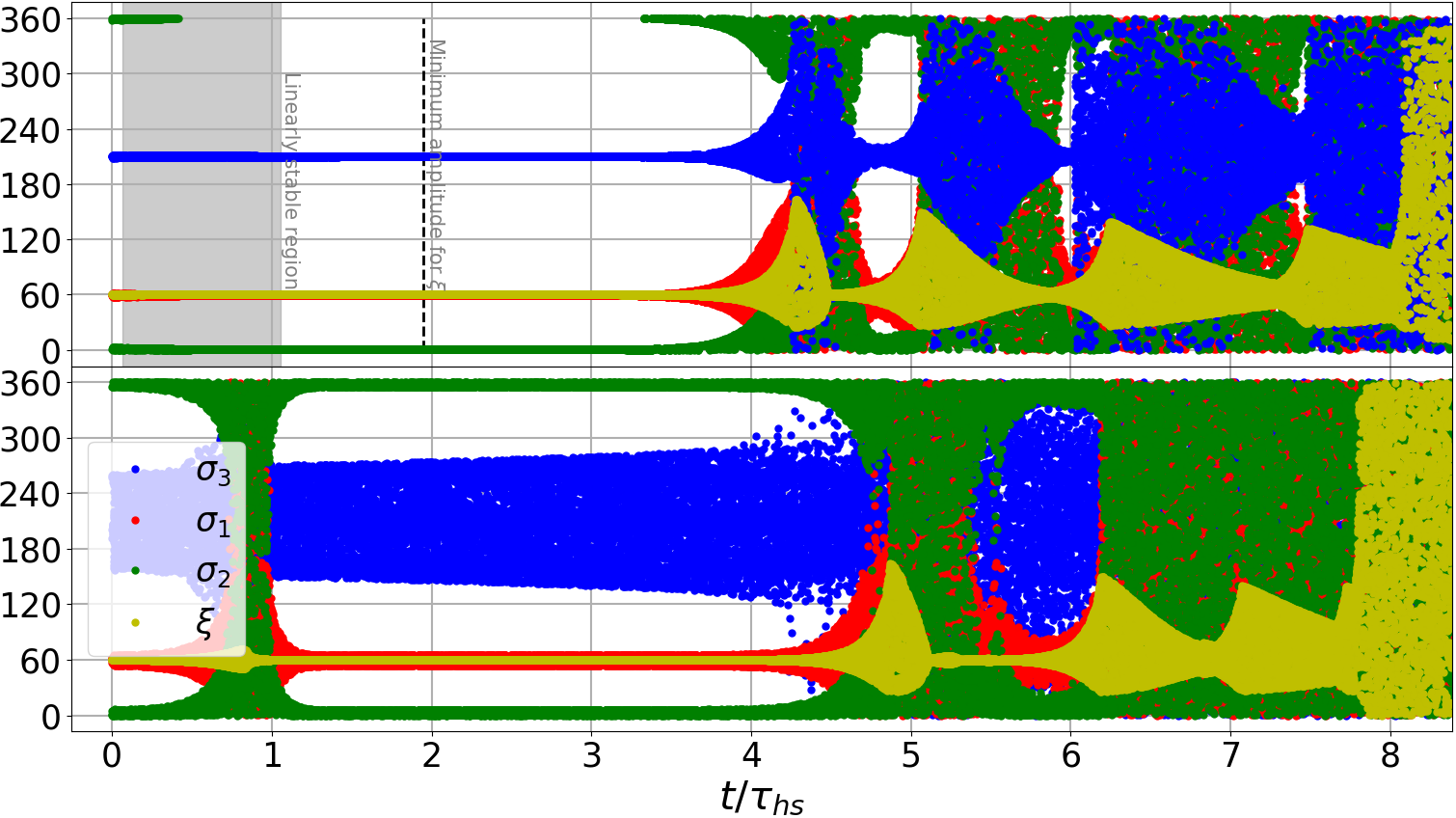}
	\caption{Evolution of $\xi$ and the $\sigma_j$ (in degrees) as a function of time for system $0$ as integrated by the simplified model $F$ (Eqs.\,(\ref{total_hamiltonian}) and (\ref{vector_field_tides})) (top) and by the direct $N$-body simulation (\ref{nbody_direct}) (bottom). In the direct simulation, the horseshoe-shaped orbits are reached after $7.8\,\tau_{\text{hs}}$, and the co-orbital motion is destroyed shortly after that, as expected from the stability map (Fig.~\ref{fig_map}). The model reaches horseshoe-shaped orbits at $8.1\,\tau_{\text{hs}}$. The negative real parts of all the eigenvalues in the grey-shaded area allow the libration amplitude of $\xi$ to reach values as small as $3.5$ arc seconds at $t=1.95\,\tau_{\text{hs}}$. In this figure as well as in Fig. \ref{fig_angle_1}, several occurrences of the \textit{eccentricity damping stabilisation} prevent the angle $\xi$ from reaching the horseshoe-shaped orbits and increase the lifetime of the co-orbital planets. The early augmentation of the libration amplitude of the angles in the bottom plot is due to the fact that in the direct simulation, the \textit{linearly stable region} is not exactly at the same values of $\delta$ as in the simplified model.} 
	\label{fig_angle_0}
\end{figure*}

When the co-orbital planets are alone, the positivity of $\R_{\text{e}}\left(\lambda\right)$ in Eq. (\ref{Re_lambda}) ensures that the system systematically reaches the horseshoe-shaped orbits and is destroyed by close encounters. In that case, the time $\tau_{\text{hs}}$ needed to reach the horseshoe-shaped orbits is \citep{CoRoCo2021}
\begin{equation}\label{tau_hs}
\tau_{\text{hs}}=\frac{\varepsilon}{9\pi\Omega}\frac{x\left(1+y\right)}{1+yx^2}\ln\left(\frac{60^{\circ}}{\Delta\xi}\right)T,
\end{equation}
where $T=2\pi/\eta$ is the co-orbital period. Denoting $\tau_{\text{dest}}$ the co-orbital lifetime without third planet, $\tau_{\text{hs}}$ does not significantly differ from $\tau_{\text{dest}}$ for a wide range of total co-orbital mass, and as long as $10^{-9}\lesssim\varepsilon\lesssim 0.005$, we have $1/2\lesssim\tau_{\text{dest}}/\tau_{\text{hs}}\lesssim 2$ \citep{CoRoCo2021}.
We can thus consider that $\tau_{\text{hs}}$ is the lifetime of the co-orbital pair, in the absence of the third planet\footnote{Especially for $\varepsilon=2\times10^{-4}$, for which $\tau_{\text{dest}}\approx1.1\,\tau_{\text{hs}}$.}. For $\Delta\xi=0.1^{\circ}$, in the case of two co-orbital Earth-like planets \citep[using the tidal parameters of][]{Lainey2016}, we have\footnote{For the constant-$Q$ model, the exponent is $6.5$ instead of $8$.}
\begin{equation}\label{tau_hs_num}
\tau_{\text{hs}}=3.771\;\text{Gyr}\;\left(\frac{\bar{a}}{0.04\;\text{AU}}\right)^8\left(\frac{m_0}{m_{\odot}}\right)^{-3/2}.
\end{equation}
We give in appendix \ref{append_tau_hs} the time $\tau_{\text{hs}}$, computed from Eq. (\ref{tau_hs}), for a variety of hypothetical co-orbital pairs.
We normalise the time by $\tau_{\text{hs}}$ in Figs. \ref{fig_angle_1}, \ref{fig_angle_0}, \ref{fig_stab_mech}, and in appendix \ref{append_simu}.

\subsection{Mechanisms of co-orbital stabilisation}
In Figs. \ref{fig_angle_1} and \ref{fig_angle_0}, the angles $\xi$ and $\sigma_j$ are plotted as a function of time. The system spends a large amount of time close to the main branch of equilibria, which allows the co-orbitals to live notably longer with the presence of the third planet. This can be seen from the destruction occuring at a time $t>\tau_{\text{hs}}$. Indeed, when the system crosses the \textit{linearly stable region}, the libration amplitude of $\xi$ decreases instead of increasing exponentially, since the real parts of all the eigenvalues of the linearised system associated to $F$ are negative. When the system leaves this region due to the drift in $\delta$ and the real part of one eigenvalue becomes positive again, the libration amplitude of $\xi$ is much smaller than it was before entering the \textit{linearly stable region}. As a result, the system needs more time to reach large libration amplitudes and settle in horseshoe-shaped orbits, which delays the co-orbital destruction. That is, crossing the \textit{linearly stable region} while being sufficiently close to the main branch (so that the linear dynamics dominates) guarantees a co-orbital lifetime longer than without the third planet.
\begin{figure*}[h]
	\centering
	\includegraphics[width=1\linewidth]{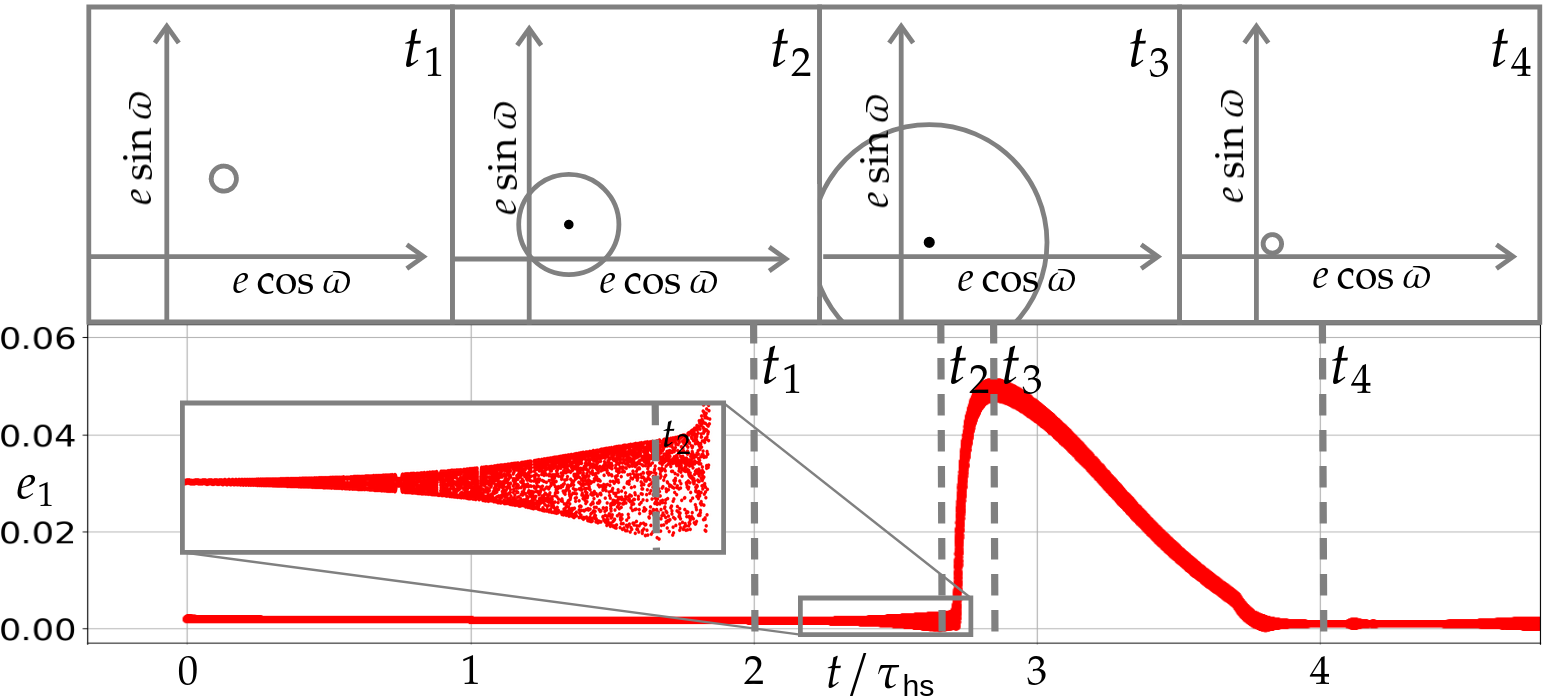}
	\caption{Value of $e_1$ as a function of time for system $1$ in Table \ref{simu_param}, as integrated by the simplified model $F$ (Eqs.\,(\ref{total_hamiltonian}) and (\ref{vector_field_tides})) (bottom), and a schematic representation of the eccentricity vector $e_1e^{i\varpi_1}$, for four particular times (top). The circle described by $e_1e^{i\varpi_1}$ grows because at least one eigenvalue's real part is positive (Fig. \ref{fig_Re}), until it surrounds the origin for $t>t_2$, which triggers a jump in eccentricity. At $t=t_3$, the eccentricity damping predicted by Eq. (\ref{tau_j}) forces the eccentricities to decrease (through non-linear contributions of $F$), and since $\xi$ is coupled with the eccentricities, it is also damped: this is the \textit{eccentricity damping stabilisation} mechanism. Fig. \ref{fig_e1_O1+reso_nu_nu3} shows that along the main branch, smaller values of $\delta$ induce smaller eccentricities, and so the drift in $\delta$ brings the center of the circle closer to the origin.}\label{fig_stab_mech}
\end{figure*}

However, the \textit{linearly stable region} is not the only reason why the co-orbitals in resonant chains can live longer. Another phenomenon, which we refer to as \textit{eccentricity damping stabilisation} in the rest of this work, allows the libration amplitude to not cross the separatrix leading to horseshoe-shaped orbits. After an exponential increase of the libration amplitude of $\xi$ and the $\sigma_j$, due to at least one eigenvalue with strictly positive real part, the amplitudes suddenly decrease and the system returns close to the equilibria. This stabilisation of $\xi$, due to eccentricity damping (see Fig. \ref{fig_stab_mech}), can happen several times before horseshoe-shaped orbits are finally reached, and the system destroyed (see Fig \ref{fig_angle_0}).
The explanation of the \textit{eccentricity damping stabilisation} relies on the behaviour of the eccentricities. In Fig. \ref{fig_stab_mech}, we plot the eccentricity $e_1$ of planet $1$ as a function of time, together with a schema of its behaviour in the plane $\left(e\cos\varpi,e\sin\varpi\right)$. At time $t=t_1$, the system is still close to the fixed points, hence the quantity $e_1e^{i\varpi_1}$ (or any of the two other eccentricities) describes a circle of small radius. Outside the \textit{linearly stable region}, the eigenvalues of the linearised systems have one positive real parts, and as time evolves, the radius of the circle grows, while its center (the equilibrium position of $e_1e^{i\sigma_1}$) gets closer to the origin due to the drift in $\delta$. At time $t>t_2$, the circle $e_1e^{i\varpi_1}$ surrounds the origin and keeps growing, as predicted by the eigenvalues, which triggers a jump in the eccentricity. On one hand, the linearised system predicts that the circle drawn by $e_1e^{i\varpi_1}$ grows to infinity, but on the other hand, tides impose an exponential decay of the eccentricities. Indeed, the first line of Eqs. (\ref{vector_field_tides}) yields \citep{Correia2009}
\begin{equation}\label{tau_j}
\dot{e}_j=-\frac{e_j}{\tau_j},\place\tau_j=\frac{2}{21}\frac{m_j}{m_0}\frac{Q_j}{q_j}n_{j,0}^{-1},
\end{equation}
and so, at time $t=t_3$, the circle reaches its maximum radius, the system is now far from its equilibrium, and tides, through non-linear contributions of the vector field $F$, force the eccentricities to decrease, which brings the system back to the vicinity of the fixed point at $t=t_4$, where the libration amplitude of the $\sigma_j$, but also of $\xi$, is small.

While the stability induced by the \textit{linearly stable region} comes from linear contributions of the vector field $F$, the \textit{eccentricity damping stabilisation} comes from non-linear contributions. This latter mechanism works thanks to a strong coupling between the eccentricities $\left(D_j,\sigma_j\right)$ and the co-orbital angle $\left(L,\xi\right)$. Indeed, in the region $\delta<0$ (tidally interesting), at most one eigenvalue has a positive real part (see Fig. \ref{fig_Re}), but due to the coupling, it allows an exponential growth of the libration angle $\xi$, as well as the eccentricities, which makes the \textit{eccentricity damping stabilisation} possible. When the time $t=t_3$ is reached, the coupling ensures that the eccentricity damping also induces a damping of the libration angle $\xi$, hence the stabilisation of the co-orbital motion. In absence of the third planet, \cite{CoRoCo2021} showed that the eccentricities are uncoupled from the co-orbital angle $\xi$. This means that the positive real part (\ref{Re_lambda}), associated to $\left(L,\xi\right)$, does not induce an exponential growth of the eccentricities, which are on the contrary damped to $0$ due to negative real parts of their eigenvalues. In this case, because of the decoupling, even if some other mechanism increases the eccentricities, the eccentricity damping predicted by Eq. (\ref{tau_j}) still occurs, but it does not induce a stabilisation of $\xi$. 

The occurrence of the \textit{eccentricity damping stabilisation} is not systematic. It occurs only if the time $t=t_3$ happens before the co-orbital planets reach horseshoe-shaped orbits. If not, the co-orbitals are destroyed before the exponential decrease of the excentricities can save them. Deciding whether or not a given system will be saved by the \textit{eccentricity damping stabilisation} requires to know the proper modes of the linearised system associated to $F$ and how $\left(L,\xi\right)$ and the $\left(D_j,\sigma_j\right)$ are written in the corresponding diagonal basis. Only a numerical work is possible, and we did not undertake it, since it is much easier to simply run the corresponding simulation.

If a larger initial $\delta$-value is chosen in these simulations, the system initially has at least one positive real part and moves away from the fixed point at exponential speed. If the \textit{eccentricity damping stabilisation} works, or if the initial $\delta$ is small enough, the \textit{linearly stable region} is reached. However, if the system reaches the \textit{linearly stable region} while being too far from the equilibria, non-linear contributions of $F$, combined with the chaotic motion induced by the $1:1$ secular resonance between $\nu$ and $\nu_3$ (see the stability map in Fig. \ref{fig_map}), can lead to peculiar orbits\footnote{e.g. switching between the Lagrangian equilibria $L_4$ and $L_5$, that is, permutation of the co-orbitals.}. Entering the \textit{linearly stable region} while still being close enough to the equilibria ensures a convergence towards the main branch, and thus an increased stability.

\subsection{Discussion}

It can be seen on Fig. \ref{fig_nu3_sur_eta} that for $\delta=-5.46$, the system is already far from the exact resonance. In fact, for system $1$ at $\delta=-5.46$, we have $n_1/n_3=2.072$. As time goes by, $\delta$ drifts towards more negative values, and at $t=6.42\,\tau_{\text{hs}}$, when it is about to reach horseshoe-shaped orbits and be destroyed, system $1$ verifies $\delta=-13.77$ and $n_1/n_3=2.186$. Similar considerations are valid for system $0$, which means that the system is already outside the resonance, but it is still influenced by the chain. As the system leaves the resonance due to the drift in $\delta$, the coupling between the eccentricities and $\xi$ becomes weaker, meaning that the \textit{eccentricity damping stabilisation} ends up failing. This prevents the co-orbitals from living forever. We nevertheless checked that it can work for simulations starting at $n_1/n_3=2.37$. Only positive values of $\delta$ allow for $n_1/n_3$ a value close to $\left(p+1\right)/p=2$ (see Fig. \ref{fig_nu3_sur_eta}).
When a positive value of $\delta$ is chosen at t = 0, the quick drift in $\delta$ due to the high values of the eccentricities (see Eq. (\ref{gamma_drift})) forces the system to reach the region $\delta<0$ in a timescale much smaller than the timescale of increase of the libration amplitude of the angles. For systems on the main branch, this means that tides favour for the ratio $n_1/n_3$ values above their Keplerian value. This result was shown by \cite{DeLaCo2014} for a two-planet chain and is confirmed by the observations of the \textit{Kepler} mission, where a large number of exoplanets were discovered with a mean motion ratio slightly larger than $\left(p+1\right)/p$ \citep[e.g.][]{Delisle_Laskar_2014}.

This section shows that our simplified model (Eqs. (\ref{total_hamiltonian}) and (\ref{vector_field_tides})) is able to satisfyingly predict the tidal evolution of a resonance chain of the form $p:p:p+1$, at least qualitatively, since a precise quantitative description can only be achieved by running the simulation of the direct set of Eqs. (\ref{nbody_direct}). This contrasts with the analytical work performed by \cite{CoRoCo2021} in the case of alone co-orbitals, where the secular model is able to quantitatively predict the outcome of the direct simulations of the complete system with less than $1$\% relative error (see their Table 3).

The influence of $m_3/\left(m_1+m_2\right)$ on the co-orbital dynamics is thoroughly discussed in appendix \ref{append_simu}. We show that, for a large $m_3$, the \textit{linearly stable region} is poorly efficient in stabilising the libration amplitude of $\xi$, while the \textit{eccentricity damping stabilisation} is very efficient (see Fig. \ref{fig_m3=8}). As $m_3$ decreases, the \textit{eccentricity damping stabilisation} loses efficiency until it does not occur anymore for very small $m_3$-values (see Fig. \ref{fig_m3=1sur32}). 
The \textit{linearly stable region} has a maximum efficiency for $m_3\approx0.29\left(m_1+m_2\right)$ (see Fig. \ref{fig_m3=0.29}). For a small value of $m_3$, it can stabilise $\xi$ only in a tiny neighbourhood around the $1\!:\!1$ secular resonance between $\nu$ and $\nu_3$ (see Fig. \ref{fig_m3=1sur32}). Finally, a premature destruction of the co-orbital motion (at $t<\tau_{\text{hs}}$) can occur for a very large $m_3$-value, if the system is already far from the resonance, at a $\delta$-value much lower than its value at the $1:1$ secular resonance between $\nu$ and $\nu_3$ (see Fig. \ref{fig_m3=19}).

In terms of co-orbital lifetime, the worst-case scenario occurs when the \textit{eccentricity damping stabilisation} does not work and when the \textit{linearly stable region} is not crossed (or is crossed while being too far from the main branch). In these cases, the only positive real part in the region $\delta<0$ has often a value close to $\R_{\text{e}}\left(\lambda\right)$ (Fig. \ref{fig_Re}), and the system reaches the horseshoe-shaped orbits in a time close to $\tau_{\text{hs}}$. 

In brief, in most cases the resonance chain increases the co-orbital lifetime, but in a few cases it can also decrease it, especially when $m_3\gg m_1+m_2$ and $\delta$ is very negative (see Fig. \ref{fig_m3=19}).

\section{Conclusion}\label{S5}

In this work, we have studied the dynamics of a pair of co-orbital planets in presence of a first-order resonance with a third planet, orbiting outside the co-orbitals. We have shown that for systems deep inside this resonance, many equilibria (or rather \textit{libration centres}) exist, at least three of them being stable. The existence of a secular resonance between the libration frequency of the co-orbitals and the precession frequency of the pericentres can lead to chaotic orbits in the conservative case. However, when tides are involved, we show that this resonance stabilises the co-orbital dynamics. 
Another stabilisation mechanism, due to eccentricity damping, is presented and explained. The model that we built is able to predict the position of the \textit{libration centres} of the complete system and we developed an algorithm to find them exactly. When tides are involved, the model reliably gives the qualitative behaviour of the system, and to a certain extent, its quantitative behaviour.

This work shows that when tidal dissipation is included, co-orbital systems are more stable if they are inside a resonance chain of the form $p:p:p+1$, which increases the chances of a still-to-come detection of a co-orbital pair of exoplanets, since \cite{Leleu2019} have shown that co-orbital pairs are often formed within a resonance chain. While the analytical work of this paper is performed for any value of the integer $p$, figures are restricted to the chain $1:1:2$ where $p=1$. We nevertheless checked that higher values of $p$ do not impact the qualitative results.

One important contribution of this work is the discovery of a $1:1$ secular resonance between the libration of the critical angle $\lambda_1-\lambda_2$ and the precession of the pericentres $\varpi_j$, as well as the inherent dynamical consequences (see Figs. \ref{fig_map} and \ref{fig_Re}). \textit{Libration centres}, quasi-periodic orbits of the unaveraged problem that generalise the equilibria of the averaged model (Sect. \ref{pf_vs_centre_libration}), are such that all the pericentres precess at the same frequency in their vicinity, which holds true to every resonance chain of any number of planets. We thus expect, for other resonance chains, the existence of similar secular resonances (e.g. between $\xi=\lambda_1-4\lambda_2+3\lambda_3$ and the $\varpi_j$ for the chain $1:2:3$).

\begin{acknowledgements}
The authors thank Jean-Baptiste Delisle and Adrien Leleu for fruitful discussions and subsequent improvements of the paper.
This work was supported by
CFisUC (UIDB/04564/2020 and UIDP/04564/2020),
GRAVITY (PTDC/FIS-AST/7002/2020),
PHOBOS (POCI-01-0145-FEDER-029932), and
ENGAGE SKA (POCI-01-0145-FEDER-022217),
funded by COMPETE 2020 and FCT, Portugal.
\end{acknowledgements}

\newpage
\bibliographystyle{apalike}
\bibliography{\bibpath biblio_nouveau_bis.bib}


\appendix
\section{Notations}\label{append_notation}

For convenience, we gather in Table \ref{notation} the notations used throughout this work\footnote{$\qoppa$ (qoppa) is an archa\"ic Greek letter.}.

\begin{table*}[h]
	\begin{center}
		\begin{tabular}{ll}
			\hline
			$m_0,\,m_1,\,m_2,\,m_3$&Masses of the star, leading and trailing co-orbitals and outermost planet\\
			$R_1,\,R_2,\,R_3$&Radii of the leading and trailing co-orbitals and the outermost planet\\
			$p$&Integer such that the resonance chain is $p:p:p+1$\\
			$\mathcal{G},\,\beta_j,\,\mu_j,\,\mu_0$&Gravitational constant, $m_0m_j/\left(m_0+m_j\right)$, $\mathcal{G}\left(m_0+m_j\right)$, $\mathcal{G}m_0$\\
			$a_j,\,e_j,\,\lambda_j,\,\varpi_j$&Semimajor axis, eccentricity, mean longitude, longitude of pericentre\\
			$\xi,\,\sigma_j,\,\Delta,\,\Lambda_j$&$\lambda_1-\lambda_2$, $-p\lambda_2+\left(p+1\right)\lambda_3-\varpi_j$, $\sqrt{2-2\cos\xi}$, see Eqs. (\ref{Lambda_def}) and (\ref{normalization})\\
			$G$ and $\Gamma,\,C_j,\,D_j$&See Eqs. (\ref{change_var_action}) and (\ref{normalization}), See Eq. (\ref{C_j}), normalised AMD (Eqs. (\ref{Lambda_def}), (\ref{change_var_action}) and (\ref{normalization}))\\
			$n_{j,0},\,a_{j,0}$&Nominal mean motions, nominal semimajor axes\\
			$\eta,\,\bar{a},\,\nu,\,\nu_3$&$n_{1,0}=n_{2,0}=\left(p+1\right)n_{3,0}/p$, $a_{1,0}=a_{2,0}$, see Eq. (\ref{nu}), see Eqs. (\ref{nu_23}) and (\ref{nu3})\\
			$C_{p,m}^{(n)},\,\delta,\,\Delta G,\,\Delta\Upsilon^{\star}$&See appendix \ref{append_coefficient}, see Eq. (\ref{delta_def}), see Eq. (\ref{Deltag_def}), see Eq. (\ref{decouplage})\\
			$\kappa_2^{(j)},\,\Delta t_j,\,\theta_j,\,\alpha_j$&Second Love number, time-lag, rotation angle, structure coefficient\\
			$Q_j,\qoppa_j,q_j,\R_j,\omega_j$&$1/\left(n_{j,0}\Delta t_j\right)$, $R_j/a_{j,0}$, $\kappa_2^{(j)}\qoppa_j^{5}$, $C_j\Lambda_j\approx1$, $\dot{\theta}_j/n_{j,0}$\\
			$F,\,F_0$&Total differential system of the model with tides, without tides\\
			$\R_{\text{e}}\left(\lambda\right),\,\tau_{\text{hs}}$&See Eq. (\ref{Re_lambda}), see Eq. (\ref{tau_hs})\\
			$\Omega,\,x,\,y,\,\varepsilon$&See Eq. (\ref{param_tides_def})\\
			\hline
		\end{tabular}
		\caption{Notations of this paper}\label{notation}
	\end{center}
\end{table*}

\section{Coefficients in the expansion of the Hamiltonian}\label{append_coefficient}

We give here the expressions of the coefficients appearing in Eqs. (\ref{O1_j3}) and (\ref{O2_j3}). They depend on the Laplace coefficients $b_{n/2}^{m}\left(\alpha\right)$ \citep{LaskarRobutel1995} and to improve readability we note $b_n^m=b_{n/2}^{m}\left(\alpha\right)$, where $\alpha=\bar{a}/a_{3,0}$. For the resonance $1:1:2$, we have
\begin{equation}
\begin{split}
&C_{1,1}^{(1)}=-\alpha b_3^1\left(\frac{7}{6}+\frac{2}{3}\alpha^{-2}+\frac{5}{3}\alpha^2\right)\\
&\place\place\;+b_3^0\left(1+\frac{5}{2}\alpha^2\right)\approx 1.1904937,\\
&C_{1,2}^{(1)}=b_3^1\left(1+\frac{3}{2}\alpha^2\right)-\frac{5}{2}\alpha b_3^0+\frac{1}{\sqrt{\alpha}}\approx -0.4283898,\\
\end{split}
\end{equation}
for the first order in eccentricity and
\begin{equation}
\begin{split}
&C_{1,1}^{(2)}=\alpha b_3^1\left(\frac{263}{168}+\frac{16}{35}\alpha^{-4}+\frac{89}{105}\alpha^{-2}+\frac{341}{105}\alpha^2+\frac{184}{35}\alpha^4\right)\!\!\!\!\!\!\!\!\!\!\!\!\!\!\!\!\!\!\\
&\place\;\;-b_3^0\left(\frac{71}{70}+\frac{24}{35}\alpha^{-2}+\frac{67}{35}\alpha^2+\frac{276}{35}\alpha^4\right)\approx -1.6957266,\!\!\!\!\!\!\!\!\!\!\!\!\!\!\!\!\!\!\\
&C_{1,2}^{(2)}=\alpha b_3^1\left(\frac{65}{24}+\frac{4}{3}\alpha^{-2}+\frac{13}{3}\alpha^2\right)\\
&\place\;\;-b_3^0\left(2+\frac{13}{2}\alpha^2\right)\approx -3.5937942,\\
&C_{1,3}^{(2)}=-b_3^1\left(\frac{29}{10}+\frac{8}{5}\alpha^{-2}+\frac{59}{10}\alpha^2+\frac{48}{5}\alpha^4\right)\\
&\place\place\;+\alpha b_3^0\left(\frac{69}{20}+\frac{12}{5}\alpha^{-2}+\frac{72}{5}\alpha^2\right)\approx 4.9668470,\\
&C_{1,4}^{(2)}=-\frac{1}{8}\alpha b_3^1\approx -0.3876274\;\;\text{ and}\\
&C_{1,5}^{(2)}=\frac{1}{2}b_3^1\left(1+\alpha^2\right)-\frac{3}{4}\alpha b_3^0\approx 0.5756950.
\end{split}
\end{equation}
for the second order.

\section{Simplified differential system}\label{append_syst_diff}

We give in this appendix the expression of the differential system derived from the Hamiltonian $\Ham_K+\Ham^{(0)}+\Ham^{(1)}$, that is, the Hamiltonian (\ref{total_hamiltonian}) truncated at first order in eccentricity, after the simplifications stated in Sect. \ref{sec_analytical} have been performed. We have
\begin{equation}
\begin{split}
&\dot{D}_1=-\frac{C_{p,1}^{(1)}\sqrt{2C_1D_1}m_1n_{3,0}}{m_0C_3}\sin\left(p\frac{\pi}{3}-\sigma_1\right),\\
&\dot{D}_2=-\frac{C_{p,1}^{(1)}\sqrt{2C_2D_2}m_2n_{3,0}}{m_0C_3}\sin\sigma_2,\\
&\dot{D}_3\!=\!-\frac{\sqrt{2C_3D_3}C_{p,2}^{(1)}n_{3,0}}{m_0C_3}\!\left(\! m_1\sin\left(p\frac{\pi}{3}\!-\!\sigma_3\right)\!-\!m_2\sin\sigma_3\right)\!,\!\!\!\!\!\!\!\!\!\!\!\!\!\!\!\!\!\!\!\!\!\!\!\!\!\!\!\\
&\dot{\Delta L}=\eta\frac{m_2\sin\xi}{m_0C_1}\left(1-\frac{1}{\Delta^3}\right),\\
&\dot{\sigma}_1=\drond{\Ham_K}{\Delta\Upsilon}+\frac{C_1C_{p,1}^{(1)}m_1n_{3,0}}{C_3\sqrt{2C_1D_1}m_0}\cos\left(p\frac{\pi}{3}-\sigma_1\right),\\
&\dot{\sigma}_2=\drond{\Ham_K}{\Delta\Upsilon}+\frac{C_2C_{p,1}^{(1)}m_2n_{3,0}}{C_3\sqrt{2C_2D_2}m_0}\cos\sigma_2,\\
&\dot{\sigma}_3\!=\!\drond{\Ham_K}{\Delta\Upsilon}\!+\!\frac{C_{p,2}^{(1)}n_{3,0}}{m_0\sqrt{2C_3D_3}}\!\left(\! m_1\cos\left(p\frac{\pi}{3}\!-\!\sigma_3\right)\!+\!m_2\cos\sigma_3\!\right),\!\!\!\!\!\!\!\!\!\!\!\!\!\!\!\!\!\!\!\!\!\!\!\!\!\!\\
&\dot{\xi}=-3\eta\left(C_1+C_2\right)\Delta L,
\end{split}
\end{equation}
where
\begin{equation}
\drond{\Ham_K}{\Delta\Upsilon}\!=\!-3\eta\left\lbrace\left(p^2C_2+p\left(p+1\right)C_3\right)\Delta\Upsilon+pC_2\Delta L^{\star}\right\rbrace.\!\!\!\!\!\!\!\!\!
\end{equation}

\section{Expression of the matrix $\mathcal{Q}_6$}\label{Q6}

We give in this appendix the matrix $\mathcal{Q}_6$ appearing in Eq. (\ref{linear_O1}). We note
\begin{equation}\label{notation_Q6}
\begin{split}
&r_j=\sqrt{2D_j},\;\;c_3=\cos\sigma_3,\;\;s_3=\sin\sigma_3,\;\;s=\sin\frac{p\pi}{3},\\&c=\cos\frac{p\pi}{3},\;\;I=3\eta p\left(C_3+p\left(C_2+C_3\right)\right),
\end{split}
\end{equation}
and obtain $\mathcal{Q}_6=$
\begin{equation*}
\left(\begin{smallmatrix}I c s {r_1}^{2} & I s r_1 r_2 & c_3 I s r_1 r_3 & I s^2 {r_1}^{2}-\nu_3 & 0 & I s_3 s r_1 r_3\\
0 & 0 & 0 & 0 & -\nu_3 & 0 \\
I s_3 c r_1 r_3 & I s_3 r_2 r_3 & c_3 I s_3 {r_3}^{2} & I s_3 s r_1 r_3 & 0 & I {s_3}^{2} {r_3}^{2}-\nu_3\!\!\! \\
\!\!\!-I c^2 {r_1}^{2}+\nu_3 & -I c r_1 r_2 & -c_3 I c r_1 r_3 & -I c s {r_1}^{2} & 0 & -I s_3 c r_1 r_3 \\
-I c r_1 r_2 & -I {r_2}^{2}+\nu_3 & -c_3 I r_2 r_3 & -I s r_1 r_2 & 0 & -I s_3 r_2 r_3 \\
-c_3 I c r_1 r_3 & -c_3 I r_2 r_3 & -{c_3}^{2} I {r_3}^{2}+\nu_3 & -c_3 I s r_1 r_3 & 0 & -c_3 I s_3 {r_3}^{2} \end{smallmatrix}\right).
\end{equation*}

\section{Direct complete model for tides}\label{append_direct}

The complete equations of motion governing the tidal evolution of a planar $(N+1)$-body system in an heliocentric reference frame, using a linear constant time-lag tidal model, are given, for $1\leq j\leq N$, by \citep{Mignard1979}

\begin{equation}\label{nbody_direct}
\begin{split}
\ddot{\vect{r}}_j 
= &-\frac{\mu_j}{r_j^3}\vect{r}_j+\sum_{k\neq j}\mathcal{G}m_k\left(\frac{\vect{r}_k-\vect{r}_j}{\left|\vect{r}_k-\vect{r}_j\right|^3}-\frac{\vect{r}_k}{r_k^3}\right)  \\&\place\place\;\;
+\frac{\vect{f}_j}{\beta_j}+\sum_{k\neq j}\frac{\vect{f}_k}{m_0} \ , \\
\ddot \theta_j 
& =\!-3\Delta t_j\frac{\kappa_2^{(j)}\mathcal{G}m_0^2R_j^3}{\alpha_jm_jr_j^8}\!\left[
\dot\theta\,r_j^2\!-\!\left(\vect{r}_j\!\times\dot{\vect{r}}_j\right)\cdot\vect{k}\right]\!,\!\!\!\!\!\!\!
\end{split}
\end{equation}
where $\vect{r}_j$ is the heliocentric position vector and $\theta_j$ the rotation angle of the planet $j$. $\vect{k}$ is the unit vector normal to the orbital plane, and $\vect{f}_j$ is the force arising from the tidal potential energy created by the deformation of planet $j$ (Eq. (\ref{total_potential}))
\begin{equation}\label{fdef}
\begin{split}
\vect{f}_j = & 
-3\frac{\kappa_2^{(j)}{\cal G}m_0^2R_j^5}{r_j^{8}}\vect{r}_j 
-3\frac{\kappa_2^{(j)}{\cal G}m_0^2R_j^5}{r_j^{10}}\Delta t_j \\&  
\left[2\left(\vect{r}_j\cdot\dot{\vect{r}}_j\right)\vect{r}_j+r_j^2\left(\dot\theta\,\vect{r}_j\times\vect{k}+\dot{\vect{r}}_j\right)\right] \ .
\end{split}
\end{equation}

\section{More numerical simulations}\label{append_simu}

In this appendix, we present the six most interesting simulations that were not shown in Sect. \ref{S4}. We particularly focus on the influence of the mass $m_3$ on the co-orbital dynamics. All the simulations comply with $m_1=m_2=10^{-4}\,m_0$, and their tidal parameters are those of system $0$ in Table \ref{simu_param}. We only integrate here the simplified model $F$ (Eq.\,(\ref{vector_field_tides})) (see Sect. \ref{sec_pseudo_fixed}). For each simulation, the real parts of the eigenvalues of the linearised system associated to $F$ are shown alongside the time evolution of the angles $\xi$ and $\sigma_j$. In the figures of the real parts, a dashed vertical black line shows the starting value of $\delta$, denoted $\delta_0$, of the corresponding simulation. In the figures of the angles, a grey-shaded area shows the \textit{linearly stable region}, when relevant. Choosing other tidal parameters does not significantly modify the figures shown here, since we normalise the real parts by $\R_{\text{e}}\left(\lambda\right)$ (see Eq. (\ref{Re_lambda})) and the times by $\tau_{\text{hs}}$ (see Eq. (\ref{tau_hs})), which is the time to reach horseshoe-shaped orbits (close to the destruction time) in the absence of a third planet. We invite the reader to have a look at Fig. \ref{fig_lsr} before reading this appendix, as $m_3$ and $\delta$ are the only varying parameters between the different simulations. For each $\delta_0$ and $m_3$, the simulation starts at the corresponding point of the main branch, with a shift $\Delta\xi=0.1^{\circ}$ to $\xi$, in order not to start exactly at the equilibrium. 
\begin{figure}[h]
	\centering
	\includegraphics[width=1\linewidth]{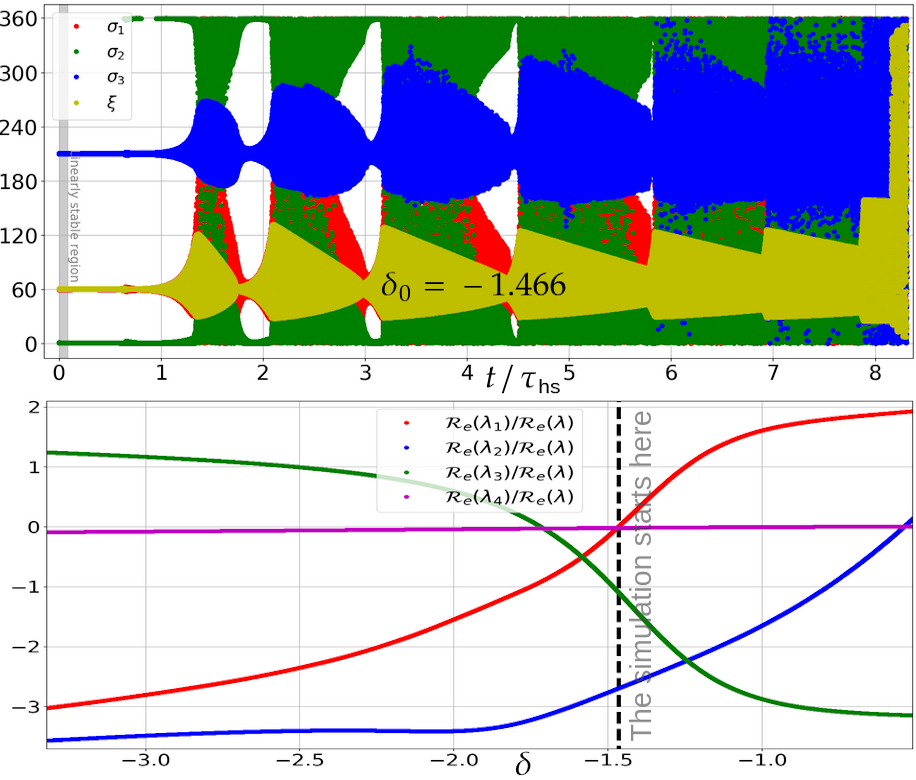}
	\caption{Here, we choose $m_3=8\left(m_1+m_2\right)$, which corresponds to the value yielding the largest \textit{linearly stable region} (see Fig. \ref{fig_lsr}). However, for this choice of $m_3$, the \textit{linearly stable region} is located at larger $\delta$-values, and so it drifts quickly (see Sect. \ref{sec_pseudo_fixed}). Furthermore, Eq. (\ref{gamma_drift}) shows that the drift in $\delta$ is proportional to $\sum_{j}m_j^{2/3}$, and here, the system leaves the \textit{linearly stable region} after less than $0.1\,\tau_{\text{hs}}$, much quicker than in Fig.~\ref{fig_m3=0.29}, where $m_3$ is smaller. The amplitude of $\xi$ reaches $0.07^{\circ}$ at its lowest, not significantly smaller than its initial value of $0.1^{\circ}$. For this choice of $m_3$, the \textit{eccentricity damping stabilisation} is very efficient though, yielding a co-orbital lifetime of $8\,\tau_{\text{hs}}$.}\label{fig_m3=8}
\end{figure}
This appendix shows that the third planet is able to decrease the lifetime of the co-orbitals only very far from the Keplerian resonance, and if $m_3\gg m_1+m_2$. For a massive third planet, the \textit{eccentricity damping stabilisation} is much more efficient in stabilising the co-orbitals than the \textit{linearly stable region}, while the \textit{linearly stable region} is more efficient for small to intermediate $m_3$-values. For a small $m_3$-value, the lifetime of the co-orbital motion is sensibly the same as in the absence of the third planet, unless the system is precisely at the $1:1$ secular resonance between $\nu$, the libration frequency of $\xi$, and $\nu_3$, the precession frequency of the pericentres, where the linear stability increases the lifetime.
\begin{figure}[h]
	\centering
	\includegraphics[width=1\linewidth]{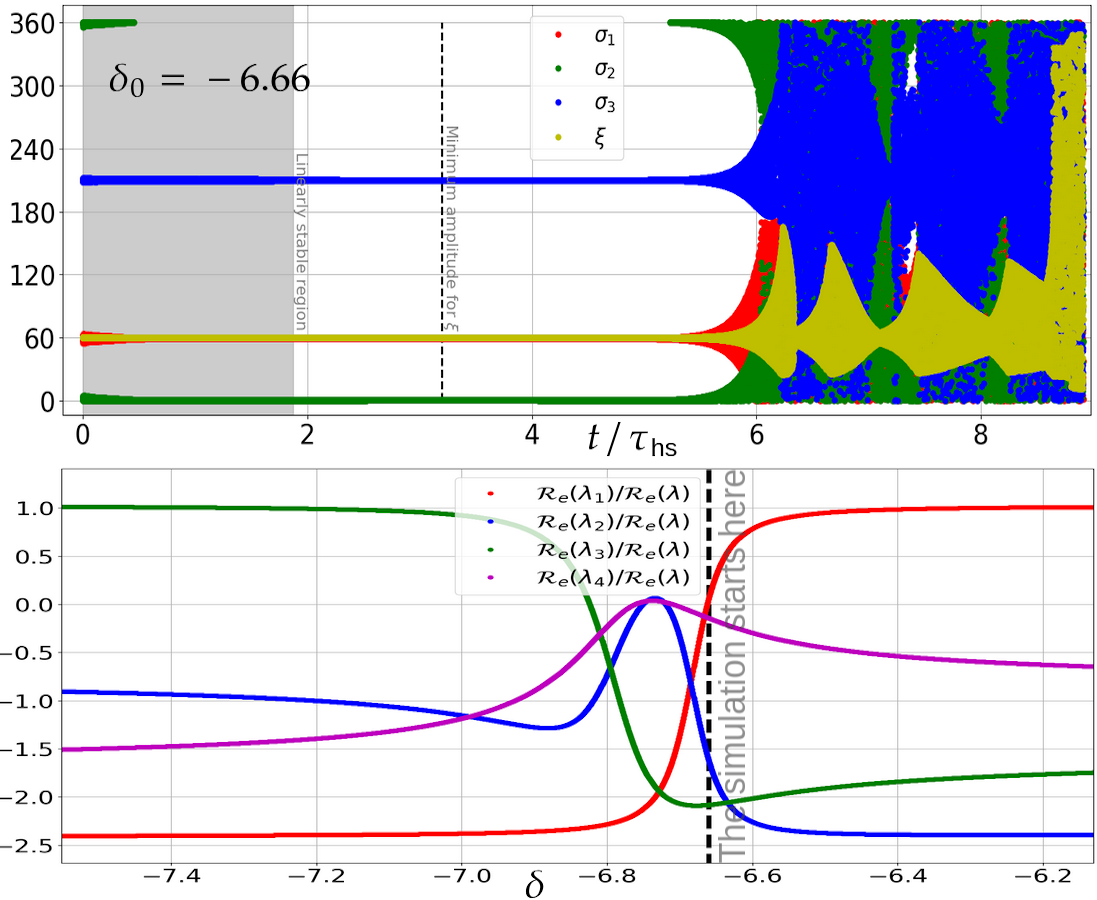}
	\caption{The best compromise between the width of the \textit{linearly stable region} and the drift in $\delta$ is for $m_3=0.29\left(m_1+m_2\right)$, which corresponds to the split of this region into two distinct strips (see Fig. \ref{fig_lsr}). Here, the system stays in this region for $2\,\tau_{\text{hs}}$, $20$ times longer than on Fig. \ref{fig_m3=8}, and at $t=3.2\,\tau_{\text{hs}}$, $\xi$ librates with only $0.0000012^{\circ}$ of amplitude, gaining a factor $80\,000$ from the initial $0.1^{\circ}$. The system stays close to the main branch for $6\,\tau_{\text{hs}}$, longer than any other simulation that we performed. The \textit{eccentricity damping stabilisation} is not as efficient as for $m_3=8\left(m_1+m_2\right)$, but still allows to reach $8.5\,\tau_{\text{hs}}$ before destruction of the co-orbital motion.}\label{fig_m3=0.29}
\end{figure}
\begin{figure}[H]
	\centering
	\includegraphics[width=1\linewidth]{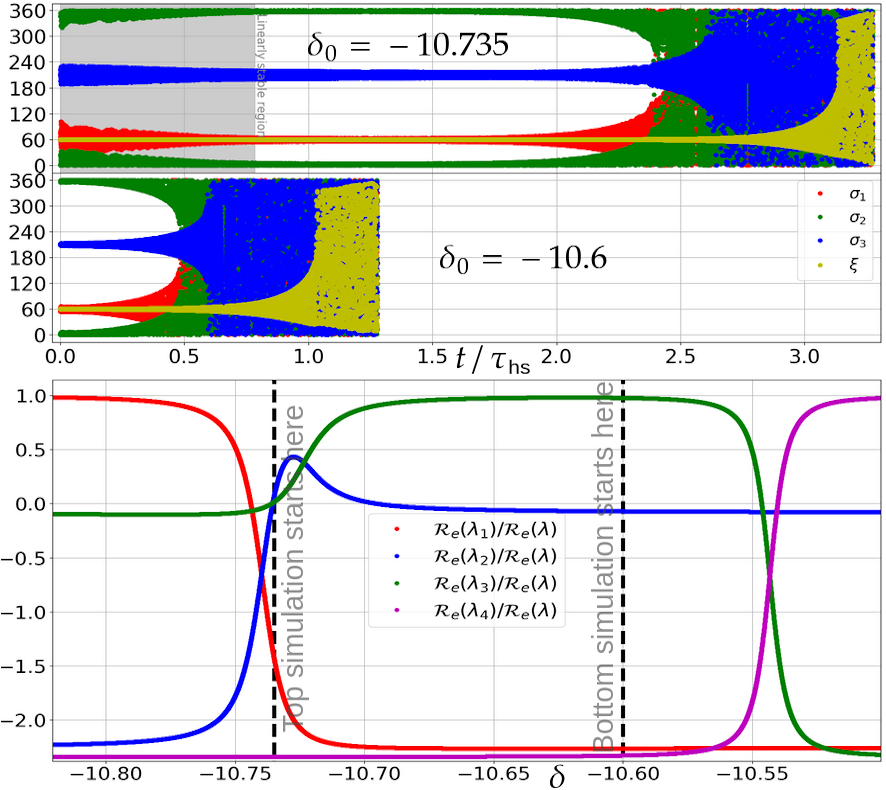}
	\caption{For $m_3=\left(m_1+m_2\right)/32$, the two \textit{linearly stable regions} are extremely narrow, and the linear stability occurs only if the system is exactly at the $1:1$ secular resonance between $\nu$ and $\nu_3$ (the initial semimajor axes between both simulations are very close with a tiny difference in $\delta$). Otherwise, the unique positive real part is $\R_{\text{e}}\left(\lambda\right)$ everywhere and the destruction occurs at $\tau_{\text{hs}}$, as in the absence of the third planet, since the \textit{eccentricity damping stabilisation} does not work.}\label{fig_m3=1sur32}
\end{figure}
\begin{figure}[h]
	\centering
	\includegraphics[width=1\linewidth]{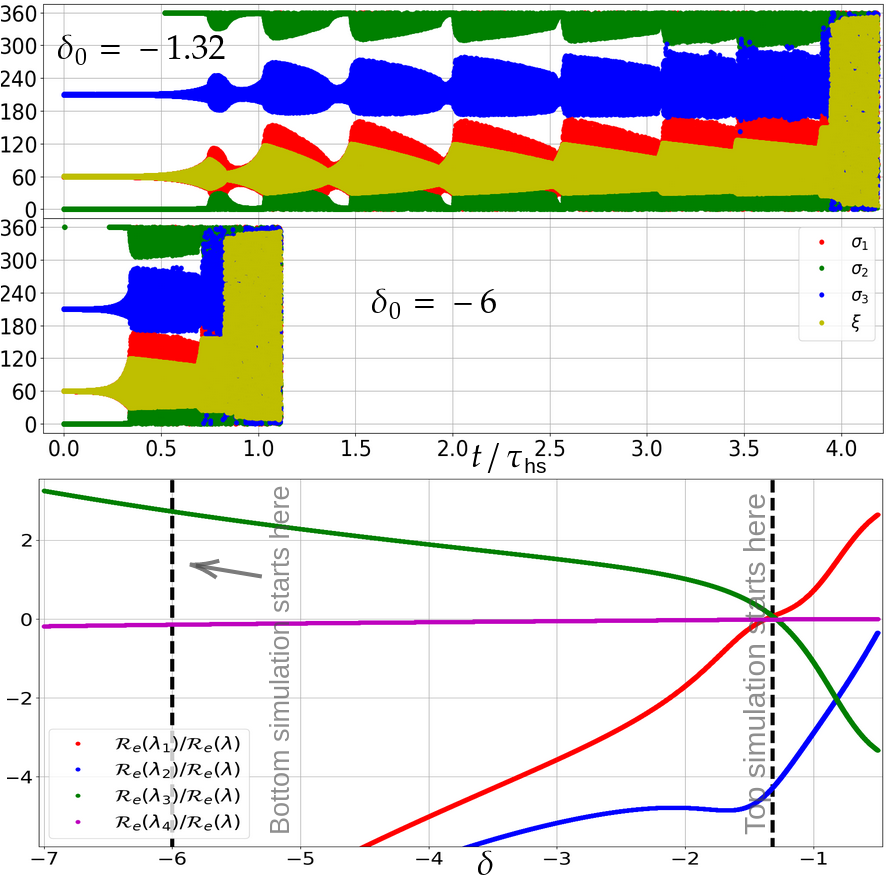}
	\caption{For $m_3=19\left(m_1+m_2\right)$, the \textit{linearly stable region} does not exist (see Fig. \ref{fig_lsr}), and we perform a simulation at the $1:1$ secular resonance between $\nu$ and $\nu_3$, where it would have been if it existed. In this region, the positive real parts are not greater than $\R_{\text{e}}\left(\lambda\right)$, and with the \textit{eccentricity damping stabilisation}, the co-orbitals almost reach $4\,\tau_{\text{hs}}$. Very far from the Keplerian resonance though ($n_1/n_3=2.6$ at $\delta=-6$, for this choice of $m_3$), the unique positive real part is significantly greater than $\R_{\text{e}}\left(\lambda\right)$. Furthermore, as $\left(L,\xi\right)$ and $\left(D_j,\sigma_j\right)$ are weakly coupled far from the Keplerian resonance, the \textit{eccentricity damping stabilisation} is poorly efficient, leading to a premature destruction (here $0.8\,\tau_{\text{hs}}$) of the co-orbital motion. Choosing a smaller $\delta$ leads to even quicker destruction.}\label{fig_m3=19}
\end{figure}

\newpage
\section{$\tau_{\text{hs}}$ for hypothetical co-orbital pairs}\label{append_tau_hs}

We give in Table \ref{tau_hs_table} the time $\tau_{\text{hs}}$ for hypothetical co-orbital pairs of exoplanets made up of solar system bodies. The semimajor axis is $\bar{a}=0.04$ AU and the mass of the host star is $m_0=m_{\odot}$, but $\tau_{\text{hs}}$ is easily deduced for other values using the exponents of Eq. (\ref{tau_hs_num}). We choose $\Delta\xi=0.1^{\circ}$, and again, it is straightforward to extend the results to another choice of $\Delta\xi$, since $\tau_{\text{hs}}\propto\ln\left(60^{\circ}/\Delta\xi\right)$ (see Eq.(\ref{tau_hs})).

\begin{table}[H]
	\begin{center}
		\begin{tabular}{lr||lr}
			\hline
			Co-orbital pair&$\tau_{\text{hs}}\!$ (Gyr)$\!$&Co-orbital pair&$\tau_{\text{hs}}\!$ (Gyr)\\
			\hline
			Earth \& Earth&$3.771$&Earth \& Moon&$44.09$\\
			Earth \& Mars&$5.480$&Earth \& Jupiter&$3.722$\\
			Earth \& Io&$2.086$&Moon \& Moon&$50.76$\\
			Moon \& Mars&$28.16$&Moon \& Jupiter&$50.63$\\
			Moon \& Io&$4.374$&Mars \& Mars&$5.761$\\
			Mars \& Jupiter&$5.747$&Mars \& Io&$2.248$\\
			Jupiter$\!$ \& $\!$Jupiter&$0.7201$&Jupiter \& Io&$2.072$\\
			Io \& Io&$2.072$&&\\
			\hline
		\end{tabular}
		\caption{$\tau_{\text{hs}}$ for some co-orbital systems.}\label{tau_hs_table}
	\end{center}
	In this table, the tidal parameters are those of \cite{Lainey2016} and only the five bodies for which $\kappa_2/Q$ is well constrained have been included. The close $\tau_{\text{hs}}$ between some systems is purely coincidental. It is due to the particular value of Jupiter's $\kappa_2/Q$ and to the fact that this body is much larger and much more massive than the other four.
\end{table}

\end{document}